\documentclass{emulateapj}
\usepackage{color}
\usepackage{graphicx}
\usepackage{dcolumn}
\usepackage{bm}
\begin{document}

\shorttitle{Depleted  cores in lenticular galaxies}
\title{Central Stellar Mass Deficits in the bulges of Local Lenticular Galaxies, and the Connection with Compact $z \sim 1.5$ Galaxies}

\shortauthors{Dullo \& Graham}
\author{Bililign T.\ Dullo\altaffilmark{1},  Alister W.\ Graham\altaffilmark{1}}
\affil{\altaffilmark{1} Centre for Astrophysics and Supercomputing, Swinburne University
of Technology, Hawthorn, Victoria 3122, Australia; Bdullo@astro.swin.edu.au}

\begin{abstract}
  We have used the full radial extent of images from the {\it Hubble Space Telescope}'s Advanced
  Camera for Surveys and Wide Field Planetary Camera 2  to extract surface brightness
  profiles from a sample of six, local
  lenticular galaxy candidates. We have modelled
  these profiles using
  a core-S\'ersic bulge {\it plus} an exponential disk model. Our lenticular disk galaxies with bulge
  magnitudes $M_{V}\la-21.30$ mag have central stellar deficits, suggesting
  that these bulges may have formed from `dry' merger events involving
  supermassive black holes while their
  surrounding disk was subsequently built up, perhaps via cold gas accretion
  scenarios. The central stellar mass deficits
  $M_{\rm def}$ are roughly 0.5 to 2 $M_{\rm BH}$ (black hole mass), rather than
  $\sim$10 to 20 $M_{\rm BH}$ as claimed from some past studies, which is in
  accord with core-S\'ersic model mass deficit measurements in elliptical
  galaxies. Furthermore, these bulges have
  S\'ersic indices n $\sim 3$, half light radii $R_{e} < 2$ kpc and masses $> 10^{11}
  M_{\sun}$, and therefore appear to be descendants of
  the compact galaxies reported at $z \sim 1.5$ to 2. Past studies which have searched for these local counterparts by using single-component galaxy models to provide the $z \sim 0$ size comparisons have over-looked these dense, compact and massive bulges in today's early-type disk galaxies. This evolutionary scenario not only accounts for what are today generally old bulges---which must be present in $z \sim 1.5$ images---residing in what are generally young disks, but it eliminates the uncomfortable suggestion of a factor of 3 to 5 growth in size for the compact, $z \sim 1.5$ galaxies that are known to possess infant disks.

\end{abstract}

\keywords{
  galaxies: elliptical and lenticular, cD -- 
 galaxies: fundamental parameter -- 
 galaxies: nuclei -- 
galaxies: photometry --
galaxies: structure  
}

\section{Introduction}
It is now widely believed that {\it all} massive elliptical galaxies, and
massive bulges in disk galaxies, house a supermassive black hole (SMBH) at
their center (Magorrian et al.\ 1998; Richstone et al.\ 1998; Ferrarese \&
Ford 2005). In the standard cosmological formation paradigm of the universe,
galaxies grow hierarchically, such that smaller systems merge to build larger
ones (e.g., White \& Rees 1978; Khochfar \& Burkert 2001).  As galaxies
containing SMBHs collide, their black holes will migrate to the center of the
merger remnant through dynamical friction, and form a bound binary. The
subsequent hardening of the black hole binary, following the dissipation of orbital 
energy and angular momentum to the nearby stars, impacts on the central stellar
distribution of the newly merged galaxy. The gravitational slingshot ejection
of these stars (from the binary's loss cone) via three-body interactions
involving the binary is thought to be responsible for the physical origin of
the partially depleted nuclear regions in luminous ``core-S\'ersic'' galaxies
which experienced `dry', i.e. gas poor, merger events (Begelman et al.\ 1980;
Ebisuzaki et al.\ 1991; Faber et al.\ 1997; Milosavljevi\'c \& Merritt 2001;
Merritt 2006; Sesana 2010; Gualandris \& Merritt 2012).

Given the above scenario, the sizes and mass deficits of partially depleted
cores are thought to reflect the amount of galactic merging and the ensuing
extent of damage caused by binary SMBHs (after having eroded any preexisting
nuclear star clusters: Bekki \& Graham 2010). Accurate measurements of the
central stellar mass deficits in ``core-S\'ersic'' galaxies can therefore
provide useful constraints on models of galaxy formation and evolution. A key
concept is that such core formation is by and large a cumulative process in
which the ejected mass from SMBH binaries scales both with the final SMBH mass
and the number of mergers (e.g.,\ Milosavljevi\'c \& Merritt 2001; Merritt 2006). Some recent studies have
additionally postulated enhanced core depletions as a result of recursive core
passages of recoiled SMBHs (Gualandris \& Merritt 2008), or due to the actions
of multiple SMBHs from merging galaxies (Kulkarni \& Loeb 2011).

To date, several quantitative studies of the central stellar mass deficit have
focused on elliptical galaxies (e.g., Graham 2004; Trujillo et al.\ 2004;
Merritt 2006; Kormendy \& Bender 2009; Dullo \& Graham 2012).  Violent, major
dry merger events are commonly, or at least traditionally, thought to produce
these elliptical galaxies (Toomre \& Toomre 1972; Barnes \& Herquist 1992;
Kauffmann \& Haehnelt 2000).  Furthermore, as noted above, dry merger events
involving supermassive black holes are also typically thought to be
responsible for producing elliptical galaxies with partially depleted cores.
Therefore, some ambiguity exists regarding the mechanisms for the formation of
the lenticular (S0) disk galaxies having depleted cores.

In 1936 Hubble introduced lenticular galaxies into the tuning fork diagram
(Jeans 1928; Hubble 1929); they were added as a hypothetical transition type
between elliptical and spiral galaxies. Later observation of the populations
of lenticular and spiral galaxies in clusters revealed the dominance of
lenticular galaxies (spiral galaxies) in local (in distant) clusters (e.g.,
Dressler 1980; Dressler et al.\ 1997; Couch et al.\ 1998; Fasano et al.\ 2000;
Desai et al.\ 2007; Poggianti et al.\ 2009; Sil'chenko et al.\ 2010).  This
suggested an evolutionary transformation of S0 galaxies from spiral galaxies,
where mechanisms such as ram pressure stripping (Gunn \& Gott 1972; Quilis et
al.\ 2000), galaxy-galaxy harassment (Moore et al.\ 1996), strangulation
(Larson et al.\ 1980; Bekki et al.\ 2002; Kawata \& Mulchaey 2008) and gas
starvation by AGN (van den bergh 2009) were forwarded to account for the
removal of disk gas and the subsequent quenching of star formation, thereby
passively fading the spiral galaxies and erasing their spiral arms. These
formation scenarios, however, do not in themselves explain the central stellar
depletions observed in some S0 galaxies.

The theoretical work of Steinmetz \& Navarro (2002) has suggested that a
galaxy's morphology is a transient phenomenon. In this hierarchical picture,
classic elliptical galaxies are built through major mergers (of disk galaxies)
and may progressively regrow stellar disks, by gas accretion, which remain
intact only until the next significant merger (see Okamoto \& Nagashima 2001;
Governato et al.\ 2009; Pichon et al. 2011; Sales et al.\ 2011; Conselice et
al.\ 2012 for supporting arguments). The number of such cycles may however be low
(i.e.\ 1) rather than several (3 to 5). Likewise, Arnold et al. (2011) and
Forbes et al.\ (2011) recently pointed out that S0s might form through a
two-phase inside-out assembly with the inner regions built early via a violent
major merger, and ``wet'' minor mergers subsequently contributing to the outer
parts.
 
This assembly scenario is consistent with the observed presence of partially
depleted cores in luminous S0 galaxies.  It also suggests that their bulges, most of 
which we know are old (e.g., MacArthur, Gonz\'alez \& Courteau 2009, and
references therein), were already around at $z \approx 1.5$ to 2. Graham
(2013, his Fig.~1) revealed that the compact galaxies at $z=1.5$ to 2 have
masses and structural properties consistent with those of the brighter bulges
in local disk galaxies. Rather than being the precursors of elliptical
galaxies prior to significant size evolution, Graham (2013) advocated that
 some of these compact high-$z$ galaxies
may therefore be associated with the bulges of modern disk galaxies. Furthermore, in this
scenario in which compact {\it galaxies} evolve into disk galaxies with
compact bulges, the (high velocity)-end of the galaxy `velocity dispersion
function' would not be expected to evolve from $z \sim 1.5$ to 0, just as observed (Bezanson,
van Dokkum \& Franx 2012).

Regarding the analysis of surface brightness profiles, Trujillo et al.\
(2004) avoided the inclusion of S0 galaxies in their sample of well-resolved
local galaxies because it was felt that the introduction of an 8-parameter
model (5 or 6 core-S\'ersic parameters for the bulge plus 2 for the
exponential disk) might be considered excessive at that time.  Ferrarese et
al.\ (2006) also avoided introducing two additional disk parameters in their
modelling of well-resolved early-type galaxies in the Virgo cluster; however
because their sample included disk galaxies, their core-S\'ersic parameters do
not correspond to the bulge component\footnote{The outer flattened disk of a
  lenticular galaxy, if not accounted for, can effectively result in a single
  `galaxy' S\'ersic index which is larger than that of the triaxial bulge
  component (e.g., Meert et al.\ 2012). Use of this larger `galaxy' S\'ersic index
  can result in an inflated measurement of the bulge's central mass deficit.}.
Although Kormendy et al.\ (2009) did fit a S\'ersic bulge plus an exponential
disk to their S0 galaxies, they marked the core region by eye rather than
objectively fitting a core-S\'ersic bulge plus exponential disk model. Depending on the sharpness of the
(inner core)-to-(outer S\'ersic) transition region, this practice may
substantially over-estimate the formal `break radius' --- which is not to be
confused with the inner or outer edge of the transition region.

Here we
endeavor to provide the most accurate measurements to date of the centrally
depleted stellar mass in lenticular disk galaxies. This is achieved by
simultaneously fitting both a core-S\'ersic model to the bulge {\it and} an
exponential model to the disk.

This paper is organized as follows. Our initial sample of six suspected
lenticular galaxies with depleted cores, plus the data reduction, and the
light profile extraction technique are discussed in Section \ref{data2}. In Dullo \&
Graham (2012) we did not have a sufficient radial extent of these six galaxies' light
profiles to include a disk component. Here we are able to test if this may
have slightly biased some of our previous measurements of the bulge
parameters. Section \ref{Modl} introduces the analytical models used to describe the
light profiles of ``core-S\'ersic'' lenticular galaxies, and Section \ref{Fita} details
our fitting analysis while in Section \ref{Cedef} we describe the method for measuring the
mass deficits of the core-S\'ersic lenticular galaxies. Our interpretations of
the central stellar mass deficits in the context of core-S\'ersic lenticular
galaxy formation scenarios are discussed in Section \ref{Dis}. In Section \ref{Fr} we discuss the formation of core-S\'ersic lenticular galaxies, and in Section \ref{hz} we
compare the physical properties of their bulges with those of compact galaxies
at $z= 1.5$ to 2. In Section \ref{6.1.2} we discuss the role of galaxy environment. Section \ref{Con2} summarizes our main conclusions. At the end of this paper are two appendices. The first provides notes on the six individual galaxies studied in this paper and the second provides a response to the criticism of Dullo \& Graham (2012) by Lauer (2012) regarding the identification of partially depleted cores. 

\section{Data and photometry}\label{data2}
\subsection{Sample selection}\label{data} 

We have targeted the six `core-S\'ersic'
lenticular\footnote{The initial morphological classification was taken from
  the Third Reference Catalog, RC3, de Vaucouleurs et al.\ (1991).} galaxies
from the sample of 39 relatively bright, nearby, early-type galaxies analysed
by Dullo \& Graham (2012). This initial sample of 39 galaxies came from Lauer
et al.\ (2005) who had claimed that they all have partially depleted cores.
Using the core-S\'ersic model rather than the Nuker model, Dullo \& Graham
(2012) subsequently revealed that seven of them did not have partially
depleted cores relative to the inward extrapolation of their outer S\'ersic
profiles (see also \ref{Sec422}). The six, suspected lenticular galaxies with
 depleted cores are listed in Table \ref{Tabb1}.

\subsection{Imaging data}
High resolution {\it Hubble Space Telescope (HST)} optical images of these
galaxies were retrieved from the public {\it HST} data archive. We only used
images from broad-band filters. While the {\it HST} Advanced Camera for Survey
(ACS; Ford et al.\ 1998) Wide Field Channel (WFC) F475W and F850LP images,
from the Virgo Cluster Survey GO-9401 program (PI: P.~C\^ot\'e), are available
for NGC 4382, for the remaining five galaxies (NGC 507, NGC 2300, NGC 3607, NGC
3706 and NGC 6849) we used the Wide Field Planetary Camera 2 (WFPC2; Holtzman
et al.\ 1995) F555W images taken from the following programs: NGC 507, NGC
3706 and NGC 6849 from the GO-6587 program (PI: D.~Richstone); NGC 2300 from
the GO-6099 program (PI: S.~Faber) and NGC 3607 from the GO-5999 program (PI:
A.~ Phillips). For NGC 3607, we also used the near-infrared NICMOS/NIC2 F160W
image observed in the GO-11219 program (PI: A.~Capetti) to enable us to better
avoid this galaxy's inner dusty spiral structure. Global properties and
observation details of all the sample galaxies are listed in
Table~\ref{Tabb1}.

The WFPC2 consists of 3 wide field cameras (WF2, WF3 and WF4) and a high
resolution Planetary Camera (PC1); each has a CCD detector with 800$\times$800
pixels. The three wide field cameras have a 0$\arcsec$.1 per pixel spatial
sampling. The smaller, high resolution planetary camera (PC1), where each
galaxy's center had been placed, has a plate scale of 0$\arcsec$.046 and a
square 36$\arcsec$$\times$36$\arcsec$ field of view (FOV).

Having two CCDs cameras with 2048 $\times$ 4096 pixels, the ACS Wide Field
Channel has a $0\arcsec.$049 pixel scale and covers a
$202\arcsec$$\times$202$\arcsec$ rhomboidal area.

Combining the constituent CCDs of each camera, the light profiles from the
WFPC2 and ACS observations probe a large range in radius, $R\ga$ 80$\arcsec$.
For comparison, Lauer et al.\ (2005, and references therein) deconvolved all
six sample galaxy images with the PSF and provided profiles out to a maximum
of $10\arcsec$. The greater radial extent that we now have, from high quality
 CCD imaging, enables us to better sample and measure the disk light.

\begin{figure}
\includegraphics[angle=270,scale=0.80]{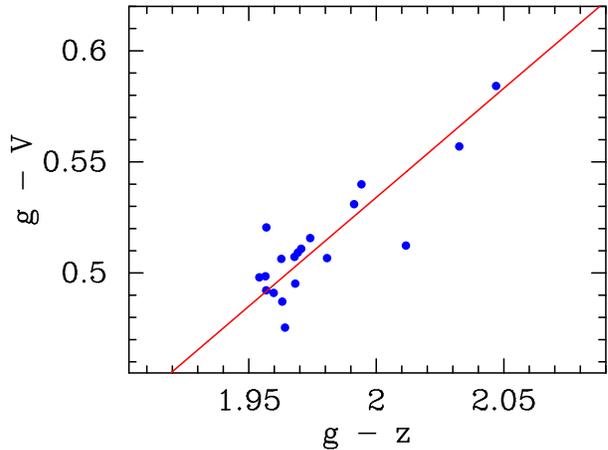}
\caption{Photometric transformation from ACS $g$ and $z$-band magnitude
  systems into WFPC2 $V$-band magnitude for NGC 4382. We use our $g$ and
  $z$-band plus Lauer et al.'s (2005) $V$-band surface brightnesses of the
  galaxy at different radii to obtain the data points.  The line is the 
least-squares
  fit to the data (Eqn.~\ref{Eqq1}). }
\label{Figg1} 
\end{figure}

While our WFPC2/F555W (roughly Jonson-Cousins $V$-band) images match Lauer et
al.'s.\ (2005) $V$-band images, for NGC 4382 the ACS/F475W (roughly SDSS $g$)
and ACS/F850LP (roughly SDSS $z$) data are transformed to the $V$-band using
 \begin{equation}
~~~~~~~~~~~~~~~~~~~~~~ g-V=0.98~(g-z)-1.43,
     \label{Eqq1}
   \end{equation}
   which is acquired here from the least squares fit to the ($g-V$) and
   ($g-z$) data shown in Fig.~\ref{Figg1}, see also Sirianni et al.\ (2005)
   and Kormendy et al.\ (2009).  The zero points in the Vega magnitude
   systems, which are adopted here to calibrate the WFPC2 (Holtzman et al.\
   1995) and ACS (Sirianni et al.\ 2005) profiles, were taken from the STScI
   web site\footnote{The ACS/WFC images used here are taken from observations
     before 2006.}.
\begin{center}
\begin{table}
\begin {minipage}{95mm}
\caption{Global properties for our sample of six suspected `core-S\'ersic' lenticular galaxies.}
\label{Tabb1}
\begin{tabular}{@{}llcccccc@{}}
\hline
\hline
Galaxy&Type& $M_{V, bulge}$ & D &$\sigma$&Filter&Exp. Time\\
&&(mag)&(Mpc)&(km s$^{-1}$)&&(s)\\
(1)&(2)&(3)&(4)&(5)&(6)&(7)\\
\multicolumn{1}{c}{} \\              
\hline                            
NGC 0507  &S0      &  $-22.56$ &$63.7^{n}$ &306&F555W&1700\\
NGC 2300   &S0              $$  &   $-21.33$ &  $25.7^{n}$&261&F555W&1520\\
NGC 3607    &S0             $$  &  $-21.55$ &$22.2^{t} $  &224&F555W&160\\
&&&&&F160W&1151\\
NGC 3706   &S0            $$  &  $-22.08$ &$45.2^{n}$ & 270&F555W&1400\\
NGC 4382  &S0                $$  &  $-21.38$& $17.9 ^{t}$  &179&F475W&750\\
                      &&&&&F850LP&1210\\
NGC 6849   &SB0           $$  &  $-22.51$ & $80.5^{n}$& 209&F555W&900\\
\hline
\end{tabular} 
Notes.---Col. (1) Galaxy name. Col. (2) Morphological type from RC3 (de
Vaucoulers et al.\ 1991). Col. (3) Absolute \emph{V}-band bulge magnitude
(galaxy magnitude for NGC 3706 since we (ultimately) adopt an elliptical morphology, see Section 4.1) obtained using our bulge-to-disk (B/D) flux ratios and absolute $V$-band {\it galaxy} magnitudes from Lauer et al.\ (2007b). These magnitudes are corrected for Galactic extinction, inclination and internal dust attenuation (Driver et al.\ 2008, their Table 1 and Eqs.~1 and 2) and $(1+z)^{4}$ surface brightness dimming, and adjusted using the distance from col. (4). Sources: ($t$) Tonry et al.\ (2001) after reducing their distance moduli by 0.06 mag (Blakeslee et al.\ 2002); ($n$) from NED (3K CMB). Col. (5) Central velocity dispersion from HyperLeda\footnote{(http://leda.univ-lyon1.fr)} (Paturel et al.\ 2003). 
\end {minipage}
\end{table}
\end{center}
    \begin{figure*}
     \includegraphics[angle=270,scale=0.38]{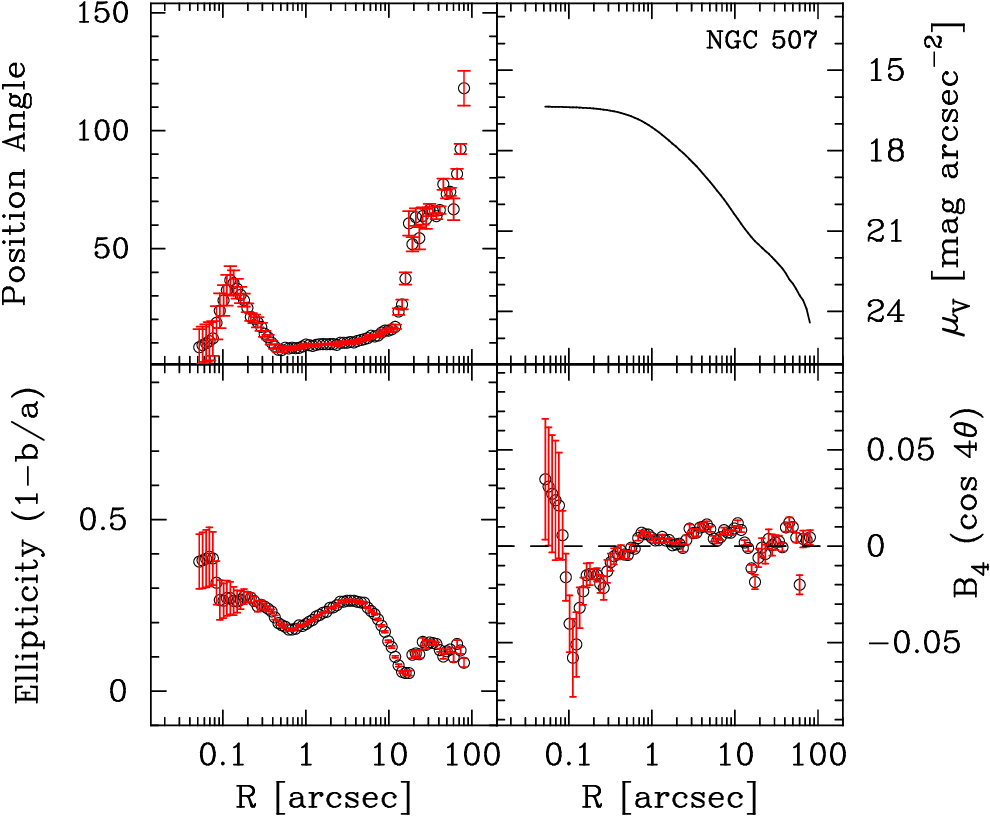}~~~~~~~~~~~~~~~
	\includegraphics[angle=270,scale=0.38]{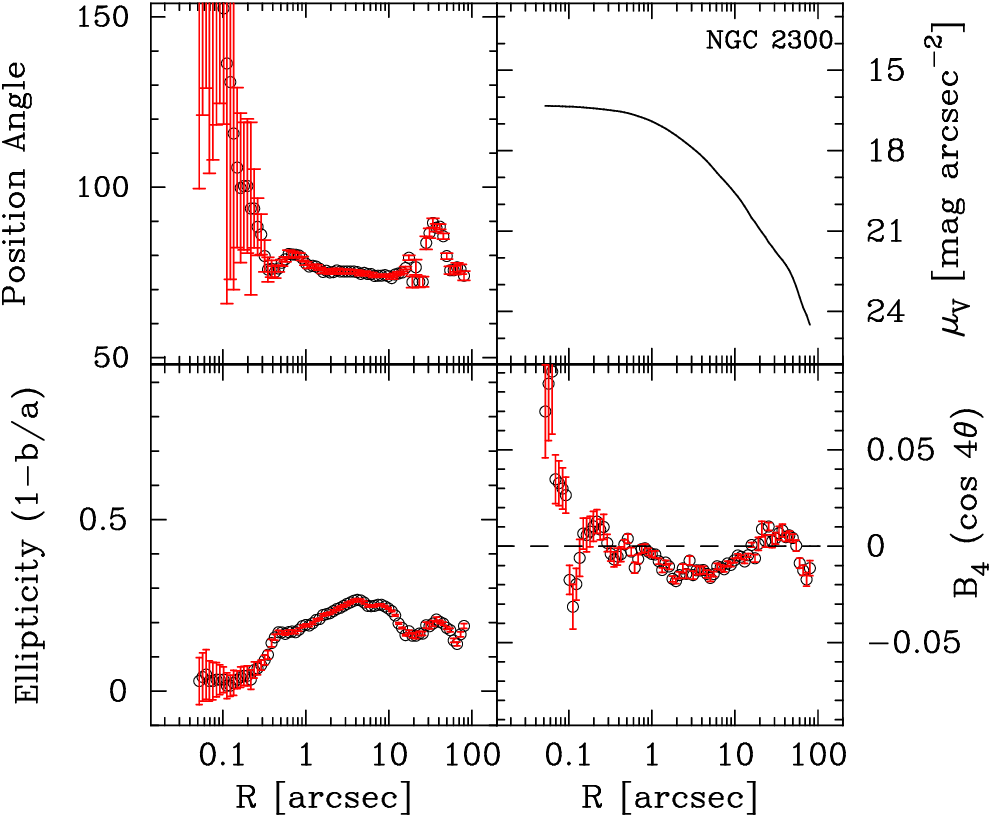}
	\end{figure*}~~~~~~~~~
	\begin{figure*}
	\includegraphics[angle=270,scale=0.38]{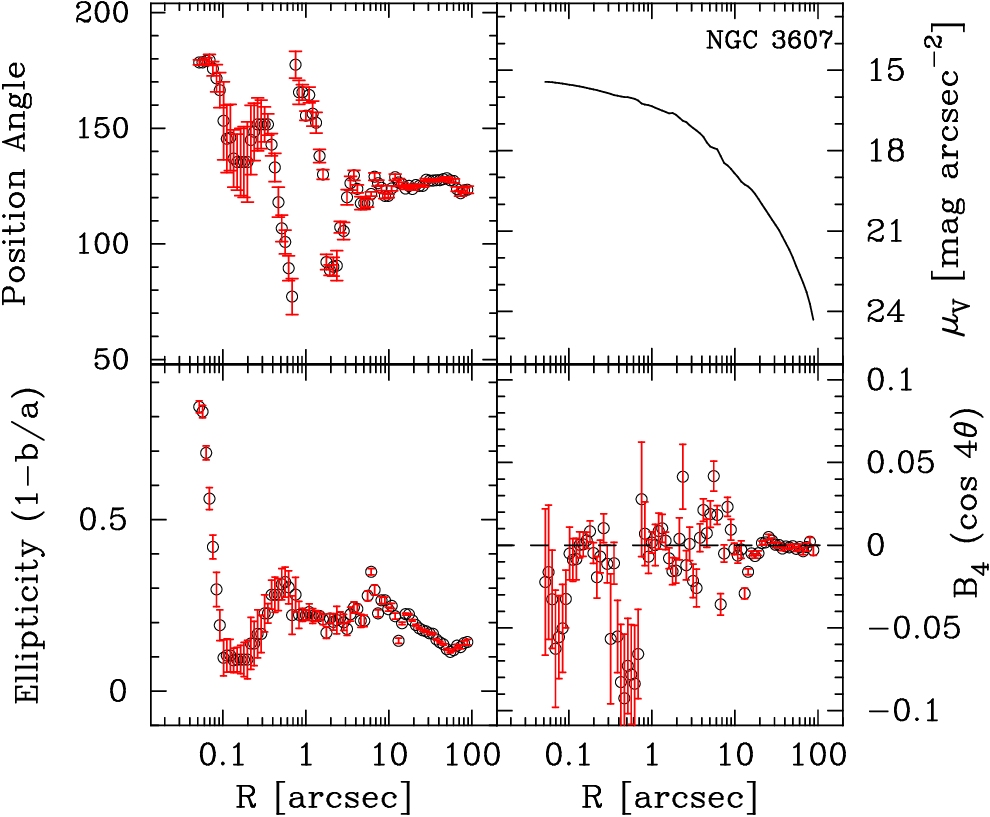}~~~~~~~~~~~~~~~
        \includegraphics[angle=270,scale=0.38]{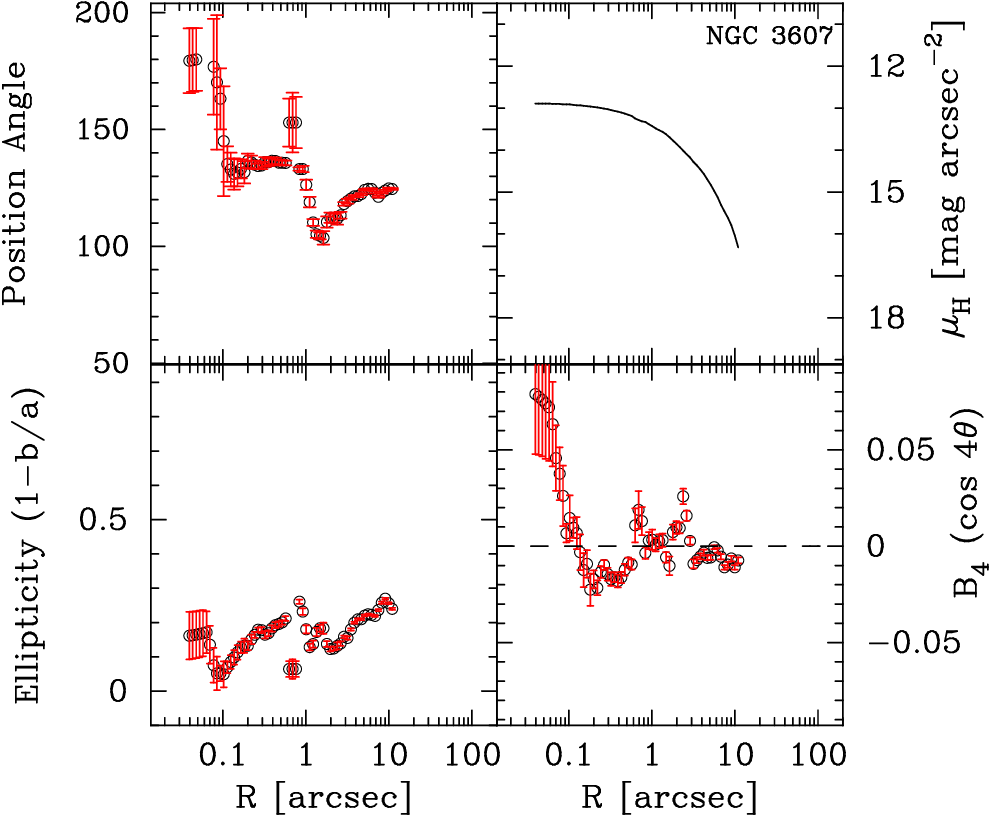}
        \end{figure*}
	\begin{figure*}
        \includegraphics[angle=270,scale=0.38]{HN3706.ps}~~~~~~~~~~~~~~~
	\includegraphics[angle=270,scale=0.38]{HN4382g.ps}

      	\end{figure*}
	\begin{figure*}
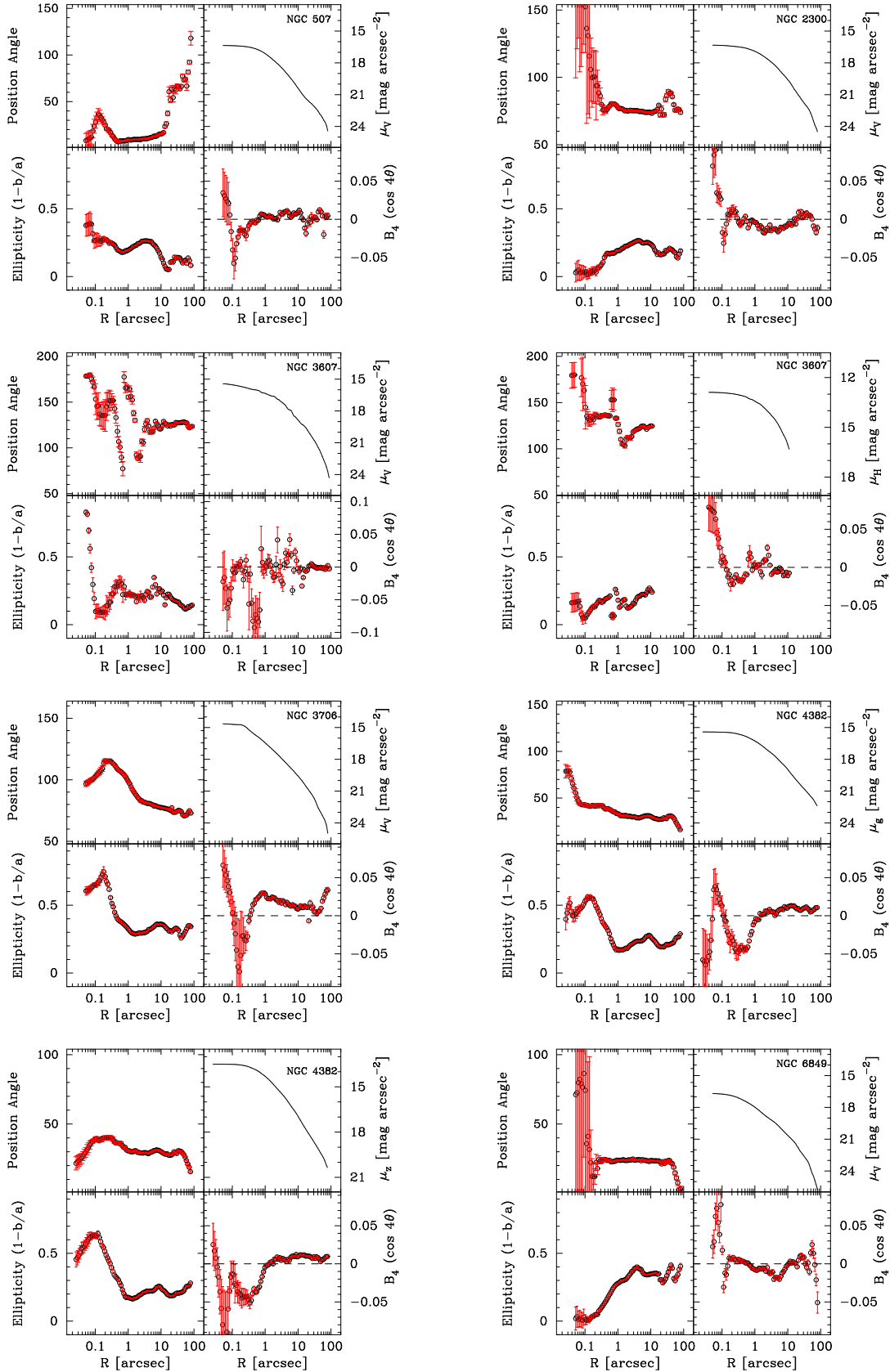

        \includegraphics[angle=270,scale=0.38]{HN4382z.ps}~~~~~~~~~~~~~~~
	\includegraphics[angle=270,scale=0.38]{HN6849.ps}
	\caption{Major-axis surface brightness, ellipticity, position angle (measured in degrees
          from north to east) and isophote shape parameter ($B_{4}$) profiles for the galaxies in
          Table~\ref{Tabb1}. Top right panels show the surface brightness
          profiles obtained with the F555W ($\sim V$-band) filter, with the exception that a NICMOS
          NIC2 F160W ({\it H}-band) profile is also shown for NGC 3607, and NGC
          4382 was imaged in the ACS F475W ($g$-band) and F850LP ($z$-band)
          filters.}
	\label{Figg2}
    	\end{figure*}
\subsection{Data Reduction}
The WFPC2, ACS and NICMOS images, retrieved from the Hubble Legacy
Archive\footnote{http://hla.stsci.edu} (HLA), were processed using the
standard HLA reduction pipeline. The reduction steps include bias subtraction,
geometric distortion correction, dark current subtraction and flat fielding
using date-appropriate references: the WFPC2 (McMaster et al.\ 2008) and ACS
(Pavlovsky et al.\ 2006) data and instrument handbooks provide detailed
descriptions of these steps.

        The automatic HLA reduction pipeline also subtracts the sky background
        values from the images. Contamination from a poor subtraction is a
        possibility which makes the automatic sky background estimation of the
        spatially extended galaxies slightly uncertain, it essentially affects
        the outermost parts (fainter regions) of such galaxies' profiles.
        Having major-axis diameters of $\la$ 4$^{\prime}$ and minor-axis
        diameters of $\la3^{\prime}$ (using $\mu_{B}$=25 mag arcsec$^{-2}$ as
        a reference level, NED), NGC 507, NGC 2300, NGC 3706 and NGC 6849 are
        within the WFPC2 field of view, thus, they are less prone to the sky
        subtraction errors. NGC 3607 and NGC 4382, however, extend beyond the
        WFPC2 and ACS field of views, respectively. Ferrarese et al.\ (2006)
        used a combination of the ACS and ground-based data to better
        constrain the sky level around NGC 4382, this data was later
        re-modelled by Dullo \& Graham (2012, their Fig.~7). We find here that
        our light profile, and fit parameters for this galaxy, match those in
        Dullo \& Graham (2012), indicating a small or zero sky subtraction
        error by the pipeline. This gives us some confidence that the
        outermost profile of NGC 3607 is also correct.

        Finally, the images were masked to avoid the gaps between individual
        CCD detectors, the partially missing quadrant of the WFPC2 images, and
        regions including bright foreground stars, background galaxies, chip
        defects and galaxy dust lanes.
\subsection{Surface photometry}\label{Sur_phot}   

The galaxy light profiles were extracted using the IRAF/STSDAS task {\sc
  ellipse} (Jedrzejewski 1987). We used the {\sc ellipse} task to construct
the best-fitting concentric elliptical isophotes to the galaxy image starting
from an initial elliptical isophote defined by the first guess values of the
isophote center (X,Y), ellipticity ($\epsilon$) and position angle (PA). We
used median filtering, logarithmic spacing, and 3$\sigma$ clipping for
flagging deviant sample points at each isophote. To extract the major-axis
light profiles, for all galaxies, the isophote centre (X ,Y), $\epsilon$ and
PA were set free to vary. For each galaxy we found that the isophote center
was stable, i.e.\ within the small quoted error box.

The deviations of isophotes from perfect ellipses can be characterized well by
higher order coefficients of the Fourier series expansion of the intensity
(Jedrzejewski 1987). In particular, the coefficient of the cos~$4\theta$ term,
$B_{4}$, is found to be of great importance in describing these deviations.
Negative B$_{4}$ values imply that the isophotes are ``boxy", but, if B$_{4}$
is positive, the isophote will be identified as ``disky". Disky isophotes are
frequently due to the presence of embedded disks in less massive, fast
rotating, dissipative (gas rich) galaxies (e.g., Carter 1978, 1987; Davies et
al.\ 1983; Bender et al.\ 1988; Peletier et al.\ 1990; Jaffe et al.\ 1994;
Faber et al.\ 1997). In contrast, dissipationless violent relaxation of stars
are often invoked to explain the boxy isophotes of triaxial,
pressure-supported massive galaxies (Nieto \& Bender 1989), which usually also
contain hot X-ray emitting gas (e.g., Cattaneo et al. 2009).

In Fig.~\ref{Figg2} we show the results of the {\sc ellipse} fitting. As noted
in Rest et al.\ (2001), the isophotal parameters derived from the {\sc
  ellipse} fits tend to be more uncertain in the very inner regions ($R\la
0.\arcsec3$); this feature is mainly due to the PSF and some contributions
from the discrete sampling and subpixel interpolation of the IRAF fitting
routine (Ravindranath et al.\ 2001). We note that while the disky/boxy
isophotal deviations vary with radius in all of the six galaxies, NGC 507 and
NGC 4382 have core regions which appear preferentially boxy within 1$\arcsec$
(Fig.~\ref{Figg2}). Nieto \& Bender (1989) reported that the S0 galaxy NGC
2300 has boxy inner ($1\arcsec < R \la 20 \arcsec$) isophotes and disky
`pointed' outer ($20\arcsec < R < 50 \arcsec$) isophotes, in agreement with
our result (Fig.\ 2).

In constructing `composite' light profiles for each galaxy, we combine the
very inner ($R \la 1\arcsec$) Lauer et al.\ (2005) PSF-deconvolved, high
resolution $V$-band light profile with our photometrically calibrated
($V$-band) profile of the PSF-unaffected region beyond $\sim$ $1\arcsec$. For
all the six galaxies in our sample, we find an excellent agreement (an overlap)
between our light profiles and those published by Lauer et al.\ (2005) over the
$ 1\arcsec - 5\arcsec$ radial range.

\section{Models For `core-S\'ersic' Lenticular galaxies} \label{Modl}
Given the two component (bulge/disk) nature of lenticular galaxies, we apply a
bulge-to-disk photometric decomposition to the one-dimensional, major-axis
surface brightness profiles of all the galaxies in our sample. When needed, a
bar component is also included.

The radial intensity distribution of our lenticular galaxies' disk component is modelled
with an exponential function given by
\begin{equation}
I_{disk}(R)=I_{0d} \exp~[-R/h],~~~~~~~~~~~~~~~~~~
\label{Eqq3}
\end{equation}
 where $I_{0d}$ and $h$ are the central intensity and scale length of the disk respectively.

We adopt the Ferrers (1877) function to describe the radial intensity distribution of the bar component, given by\\
$I_{bar}(R)=I_{0bar}\left[1-(R/a_{bar})^{2}\right]^{n_{bar}+0.5},~~~~~~~~R< a_{bar},$

\begin{equation}
I_{bar}(R)=0,~~~~~~~~~~~~~~~~~~~~~~~~~~~~~~~~~~~~~~~~~~~~~~~~~~~~~~~~~~~~~~~~~~~~ R> a_{bar},~~~~~
\label{Eqq3}
\end{equation}
where $I_{0bar}$, $a_{bar}$ and $n_{bar}$ are the central intensity, major-axis length and shape parameter of the bar, respectively (cf. Laurikainen et al.\ 2010). 

Since the work by Caon et al.\ (1993) and Andredakis, Peletier \& Balcells
(1995), several studies revealed that, in general, the S\'ersic (1963) model
describes the underlying light distributions of both elliptical galaxies and
the bulges of disk galaxies exceedingly well over a large radial range. This model can
be written as
\begin{equation}
I(R) = I_{0} \exp \left[ - b_{n}
\left(\frac{R}{R_{e}}\right)^{1/n}\right],
 \label{Eqq2}
\end{equation}
where $ I_{0}=I(R=0)$ is the central intensity. The quantity $b_{n}\approx 2n- 1/3$, for $1\la n\la 10$ (e.g.,\ Caon et al.\ 1993; Graham 2001) is a function of
the S\'ersic index $n$, and is defined in such a way to ensure that the half
light radius, $R_{e}$, encloses half of the total luminosity. Reviewed in Graham
\& Driver (2005), the total luminosity of the S\'ersic model within any radius
$R$ is given by
\begin{equation}
L_{T,Ser}(<R) = I_{e} R^{2}_{e} 2 \pi n \frac{e^{b_{n}}}{(b_{n})^{2n}} \gamma (2n,x),
 \label{Eqq2a}
\end{equation}
where $\gamma (2n,x)$ is the incomplete gamma function and $x= b_{n}(R/R_{e})^{1/n}$.

Systematic downward departures of the inner light profile relative to the
inward extrapolation of the outer S\'ersic model are known to exist in
luminous galaxy/bulge light profiles. These are not exactly the same objects as ``core'' galaxies identified with the Nuker model (Lauer et al.\ 1995; Byun et al.\ 1996) or the double power-law model of Ferrarese et al.\ (1994). ``Core-S\'ersic'' galaxies have partially-depleted cores relative to their outer S\'ersic profile whereas ``core'' galaxies need not have any deficit but instead simply an inner power-law slope $\gamma < 0.3$. As detailed in Dullo \& Graham (2012), the
core-S\'ersic model (Graham et al.\ 2003), a combination of an inner power-law
and an outer S\'ersic function, provides a good representation of the
brightness profiles of bulges in disk galaxies with depleted cores. The model
is defined as
 \begin{equation}
I(R) =I' \left[1+\left(\frac{R_{b}}{R}\right)^{\alpha}\right]^{\gamma /\alpha}
\exp \left[-b_{n}\left(\frac{R^{\alpha}+R^{\alpha}_{b}}{R_{e}^{\alpha}}
\right)^{1/(\alpha n)}\right],~~~~~ 
\label{Eqq4}
 \end{equation}
with 
\begin{equation}
I^{\prime} = I_{b}2^{-\gamma /\alpha} \exp 
\left[b_{n} (2^{1/\alpha } R_{b}/R_{e})^{1/n}\right].~~~~~~~~~~~~~~~~~~~~~~~~~~~~~~~~~~~~~~~~~~~~
\label{Eqq5}
\end{equation}
The term $I_{b}$ denotes the intensity at the core's break radius $ R_{b}$,
$\gamma$ is the slope of the inner power law, and $\alpha$ controls the
sharpness of the transition between the inner power-law and the outer S\'ersic
profile.  Both $R_{e}$ and $b_{n}$ have the same general meaning as in the
S\'ersic model.

\begin{table*}
\begin {minipage}{177mm}
~~~~~~~~~~~\caption{Structural parameters.}
\label{Tabb2}
\begin{tabular}{@{}llccccccccccccccc@{}}
\hline
\hline
Galaxy&Type&$ \mu_{b, V} $ & $R _{b}$ &$R_{b}$ &$ \gamma$&$\alpha$&$n$&$R_{e}$&$R_{e}$&$\mu_{0d, V}$&$h$&$(B/T)_{Obs}$&$ (B/T)_{Cor}$&$i$\\
&&&(arcsec)&(pc)&&&&(arcsec)&kpc&&(arcsec)&&&($^{\circ}$)\\
(1)&(2)&(3)&(4)&(5)&(6)&(7)&(8)&(9)&(10)&(11)&(12)&(13)&(14)&(15)&\\
\multicolumn{6}{c}{} \\ 
\hline                                           
NGC 0507 &S0  &16.38     &  0.33  &  102&  $ 0.07$ &5&2.19&5.34&1.65&21.03& 27.69&0.32&0.43& 0 \\
NGC 2300 &S0    &16.61 &  0.53&70&   $ 0.08$ &2&2.20&7.69&1.02&20.39&21.08&0.42&0.57&44 \\
NGC 3607 &S0&--&--&--&--&--&2.39&7.75&0.84&19.16&18.61&0.40&0.57&59\\
NGC 3706$^{\dagger}$&E    &14.16&0.11 &24& -0.02&10&6.36&42.08&9.18& -- &--&--&--&53\\
NGC 4382 &S0   &15.01      &  0.27 & 24 & 0.07   & 5&     2.65&11.14&0.99&19.50&35.07&0.28&0.40&39 \\
NGC 6849$^{\ddagger}$&SB0  &16.67& 0.18  &  69& 0.20   &5&     3.23&7.77&2.98 & 20.72&16.93&0.31&0.46&55\\
\hline
\end{tabular} 

Notes.---Structural parameters from fits to the $V$-band major-axis surface
brightness profiles (Fig.~\ref{Figg3}). 
Col. (1) Galaxy name. Col. (2) Our adopted morphological classification. Col. (3)-(10) Best-fit bulge structural parameters from the core-S\'ersic model,
Eq.~\ref{Eqq4}. Col. (11)-(12) Best-fit disk structural parameters from the exponential model,
Eq.~\ref{Eqq3}. Break surface brightness $\mu_{b,V}$ and disk central surface brightness  $\mu_{0d,V}$  are in mag arcsec$^{-2}$. Col. (13) Expected V-band bulge-to-total (B/T)$_{Obs}$ flux ratios obtained using Graham \& Driver (2005, their Eq.~19). These ratios are corrected for Galactic extinction, surface brightness dimming, internal dust attenuation and inclination (Driver et al. 2008, their Table 1 and Eqs. 1 and 2) and listed in Col. (14). Col. (15) Disk inclination angles ($i$) are derived using the galaxies' major- and minor-axis diameters from NED. $\dagger$ We classify NGC 3706 as an elliptical galaxy based on our light profile analysis. $\ddagger$ The fit parameters for the bar component: $\mu _{0bar,V}$=21.63 mag arcsec$^{-2}$, $a_{bar}=29$$\arcsec$.04, and $n_{bar}=6.37$.  
\end{minipage}
\end{table*}

\begin{figure*}
\begin {minipage}{182mm}
\includegraphics[angle=270,scale=0.76]{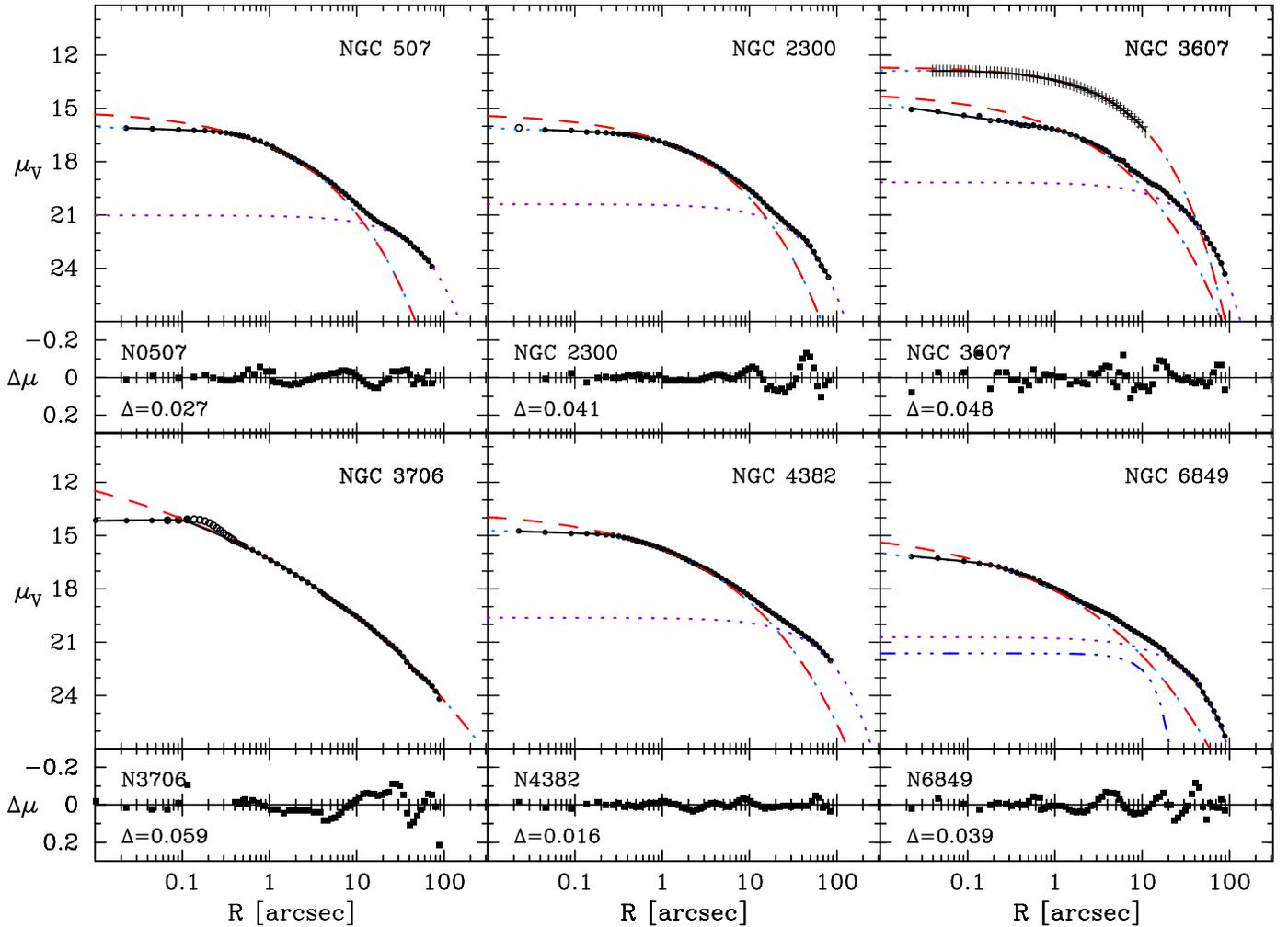}
\caption{Fits to the major-axis surface brightness profiles of the galaxies in
  Table~\ref{Tabb1}. For NGC 3607 (with a dusty nuclear spiral), the
  core-S\'ersic model fit to the PSF-affected {\it H}-band data (crosses) is
  shown in addition to the core-S\'ersic+exponential fit to the {\it V}-band
  data (dots). The dashed curves show the S\'ersic component of the
  core-S\'ersic fits to the data, while the dotted curves show the exponential
  function simultaneously fit to the large scale disks. The dash-dot-dot-dot curve shows
  the Ferrers function fit to the bar component of NGC~6849. The solid curves
  show the complete fit to the profiles, with the rms residuals, $\Delta$,
  about each fit given in the lower panels. For NGC 3607, we only show the rms
  residual from the {\it V}-band profile fit, while $\Delta$ from its {\it
    H}-band profile fit is 0.014 mag arcsec$^{-2}$.  Data points excluded from
  the fits are shown by the open circles (see Section \ref{App2}).}
\label{Figg3}
\end{minipage}
\end{figure*}

\section{Fitting analysis}\label{Fita}
Fig.~\ref{Figg3} displays the best model fits to the major-axis surface
brightness profiles of the 6 galaxies listed in Table~\ref{Tabb1}. We follow
the iterative fitting technique of Dullo \& Graham (2012), minimizing the root
mean square (rms) residuals to determine the best fit parameters that match
the data. It is important to note that these profiles now cover a large range in
radius ($ \mbox{out to}~ R\ga80\arcsec$), giving both the core-S\'ersic and
the exponential models enough radial expanse to both quantify the curvature in the
bulge profile (i.e. the S\'ersic index) and define the disk scale length.
Table~\ref{Tabb2} lists the fit parameters obtained from the adopted
core-S\'ersic plus exponential bulge+disk models.

Apparent from Fig.~\ref{Figg3} is that the global light distribution of all
the sample galaxies, except for NGC 3706 (E not S0) and NGC 6849 (barred), can
be accurately represented by the core-S\'ersic+exponential model with a small
rms residuals $\la 0.05$ mag arcsec$^{-2}$. If a single core-S\'ersic model is fit to the light profile of a disk galaxy
(i.e. a two-component system comprised of a flattened disk and a triaxial bulge), this would result in parameters that do not describe the triaxial bulge component of the galaxy, effecting in 
particular the S\'ersic index, and ultimately the mass deficit
measurement as happened in Ferrarese et al. (2006, See also Meert et
al. 2012).
Although NGC 3706 is classified as a lenticular galaxy in the RC3, we find
that its light profile is best fitted by a single core-S\'ersic model, without
an exponential disk component. This suggests that the galaxy may be an elliptical
galaxy misclassified as an S0 in the RC3, consistent with the conclusion of
Laurikainen et al.\ (2010). As for
NGC 6849, the 3-component bulge-bar-disk light profile is well described by
the core-S\'ersic bulge+Ferrers bar+exponential disk decomposition model (see
Fig.\ref{Figg3} and Section \ref{Sec416}).

While the $V$-band surface brightness profile of NGC 3607 is well described by
the core-S\'ersic bulge+exponential disk model, the 1$\arcsec$.31 core radius
of this fit is suspiciously larger than the 0$\arcsec$.22 core radius reported by Richings et al.\ (2011) from their
$H$-band profile analysis. The explanation is that a dusty nuclear spiral has
caused a reduction to the inner $V$-band light profile. Our analysis of the
PSF-affected $H$-band light profile (Fig.~\ref{Figg3}) gives a core radius of
0$\arcsec$.11, which is in fair agreement with the result found by Richings
et al.\ (2011) but does not support a large core in this galaxy.
Given that the core size we find with the PSF-affected
$H$-band data is comparable to the seeing, coupled with the dusty spiral structure,
 we are unable to conclude if there
is indeed any partially-depleted core in this galaxy, and as such we exclude
it from our final analysis.

We have found that the bulge model parameters (Table \ref{Tabb2}) generally
agree well with those from Dullo \& Graham (2012). 
For NGC 2300, the contribution of the disk light to the inner $15 \arcsec$
profile which we modelled in Dullo \& Graham (2012) did however result in a
break radius $R_{b}=0\arcsec.98$ which is slightly larger than the one from
this work ($R_{b}=0\arcsec.53$). With the exception of this galaxy, the
agreement between the break radii from these two studies is good, constrained to 
less than a 20\% discrepancy. For each
galaxy, we find that the $V$-band surface brightness at the break radius
($\mu_{b,V}$) given in Table 2 agrees with those presented in Dullo \&
Graham (2012) to within the range of error bar: there is $\sim 5\%$
discrepancy, which is well within the Dullo \& Graham (2012) $\sim 10\%$
uncertainty range. We have also compared the inner power-law slopes ($\gamma$)
from the two studies. We find that the new $\gamma$ values (Table 2) are
consistent with those reported in Dullo \& Graham (2012).
  
The five core-S\'ersic lenticular galaxies (Table \ref{Tabb2}) have bulge
S\'ersic indices $n\sim3$, while typically for core-S\'ersic elliptical
galaxies $n$ is greater than 3. This may be because, for a given magnitude,
the bulges of disk galaxies appear to have lower S\'ersic indices than
elliptical galaxies of the same magnitude (Graham 2001, his Fig.~14). It may
also be that the elliptical core-S\'ersic galaxies from Dullo \& Graham (2012)
are brighter.

\section{Central stellar mass deficit}\label{Cedef}
The central stellar mass deficit ($M_{\rm def}$) of `core-S\'ersic' galaxies
is predicted to be generated mainly via the three-body encounters of the core
stars with the inspiraling black holes in a merger remnant (e.g., Ebisuzaki et
al.\ 1991; Milosavljevi\'c \& Merritt 2001; Merritt 2006). Graham (2004)
measured this stellar mass deficit from the luminosity difference $L_{\rm
  def}$ between the inward extrapolation of the outer S\'ersic model and the
(sharp-transition\footnote{Trujillo et al.\ (2004) set the transition
  parameter $\alpha \rightarrow \infty$ to obtain the 5-parameter
  sharp-transition core-S\'ersic function given by their Eqn. 5.})
core-S\'ersic model. This approach was adopted in subsequent works by
Ferrarese et al.\ (2006), Merritt (2006) and Hyde et al.\ (2008). We adopt a
slightly different methodology to those studies by using a finite value for
$\alpha$ in Eq.~\ref{Eqq4} and thus a smoother transition region between the inner power-law and the outer
S\'ersic profile of the core-S\'ersic model. The total core-S\'ersic model
luminosity (Trujillo et al.\ 2004; their Eq.~A19) is given by

$L_{T,cS}=2\pi I'n(R_{e}/b^{n})^{2}\int\limits_{b(R_{b}/R_{e})^{1/n}}^{+\infty}e^{-x}x^{n(\gamma+\alpha)-1}$

\begin{equation}
~~~~~~~~~~~~~~~~~~~~~~~~~~~~~~~~~\times\left[x^{n\alpha}-(b^{n}R_{b}/R_{e})^{\alpha}\right]^{(2-\gamma-\alpha)/\alpha}dx,~~~~~~~
\label{Eqq6}
 \end{equation}
 where all the parameters have the same meaning as in Eq.~\ref{Eqq4}. The
 difference in luminosity between the outer S\'ersic model (Eq.~\ref{Eqq2a}) and the
 core-S\'ersic model (Eq.~\ref{Eqq6}) is of course the central stellar flux
 deficit.

 To convert the luminosity deficits into mass deficits, individual stellar
 mass-to-light ($M/L$) ratios were derived for each galaxy. To determine these
 ratios, we have made use of the available (preferentially nuclear) colours of
 the galaxies as well as the colour-age-metallicity-($M/L$) diagram from
 Graham \& Spitler (2009; their Fig~A1). As in Graham (2004), we have assumed
 that the inner regions of these galaxies have an evolved (old) single
 population of stars.

\begin{figure}
\includegraphics[angle=270,scale=0.45]{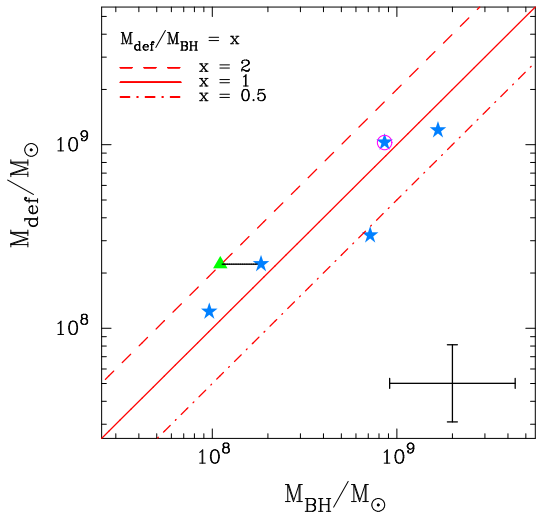}
\caption{Central mass deficit $(M_{\rm def})$ versus black hole mass
  ($M_{BH}$) for `core-S\'ersic' galaxies listed in Table~\ref{Tabb2} and
  \ref{Tabb3}. The only elliptical galaxy is circled. The $M$-$\sigma$
  relation presented in Graham et al.\ (2011) was used for estimating the SMBH
  masses of the galaxies. A horizontal solid line connects the two predicted
  black hole masses of NGC 6849: i) assuming it is a barred galaxy (triangle)
  and ii) considering it as a non-barred galaxy (star). A representative error
  bar is shown at the bottom of the panel.}
\label{Figg4} 
\end{figure}

 For NGC 2300, NGC 3607 and NGC 4382, the nuclear $V-I$ colours of 1.33, 1.36
 and 1.11, taken from Lauer et al.\ (2005), correspond to stellar $M/L_{V}$
 ratios of 5.0, 5.7 and 2.6, respectively. Using the HyperLeda database, NGC
 507 and NGC 3706 have $V-I$ colours of 1.40 and 1.34, respectively, while NGC
 6849 has $V-I$=1.03. From this, we obtain $M/L_{V}$ ratios of 5.5, 5.2, and
 2.4 for NGC 507, NGC 3706 and NGC 6849, respectively (see Table \ref{Tabb3}).
 These mass-to-light ratios were used to convert the stellar flux deficits
 into stellar mass deficits. These have been plotted in Fig.~\ref{Figg4} against each galaxy's
 expected black hole mass as derived from the $M-\sigma$ relation (Graham et
 al.\ 2011; the lower half of their Table 2) based on the $\sigma$ values in Table \ref{Tabb1}.

As noted above, high-accuracy N-body simulations by Merritt (2006) revealed that the effect of
multiple dissipationless mergers on core formation is cumulative, thus, the
extent of core evacuation could reflect the amount of merging. He found that
the total mass deficit after $N$ `dry' major mergers is $\approx$ 0.5$NM_{\rm
  BH}$, where $M_{\rm BH}$ is the final black hole mass. This result hinges on
a key assumption that the central black hole would tidally disrupt infalling
high-density satellites, thereby protecting any preexisting core (Faber
et al.\ 1997; Boylan-Kolchin \& Ma 2007).

Fig.~\ref{Figg4} shows that the mass deficits for the `core-S\'ersic'
lenticular galaxies (plus one elliptical galaxy) listed in Table \ref{Tabb2} are $ M_{\rm def}\sim
0.5-2$$M_{\rm BH}$, which translates to a few (1 to 4) `dry' major merger
events. For reference, measurements of close galaxy pairs in real data (i.e.
not simulations) suggest that massive galaxies ($> 10^{10.5} M_{\odot}$) have
experienced 0.5 to 2 major mergers since $z \sim 0.7$ to 1 (e.g., Bell et al.\
2004, 2006; Bluck et al.\ 2012; Man et al.\ 2012; Xu et al.\ 2012). Our figure
is double this value, but it has no upper redshift constraint associated with it.
The mean elliptical galaxy $M_{\rm def}/M_{\rm BH}$ mass ratio from Graham
(2004) is $2.1\pm1.1$, while the mean ratio from Ferrarese et al.\ (2006) is
$2.4\pm0.8$ after excluding the S0 galaxy NGC~4382 from their sample (as they did at
the end of their Section 5.2). We used the $M_{\rm BH}$-$\sigma$ relation presented in Graham et al.\ (2011; the lower half of their Table 2) for estimating the SMBH masses of the five core-S\'ersic galaxies shown in Table \ref{Tabb3}. While this $M_{\rm BH}$-$\sigma$ relation was constructed by combining core-S\'ersic and S\'ersic galaxies, Graham \& Scott (2012, their Table 3) have recently shown that core-S\'ersic and S\'ersic galaxies follow similar $M_{\rm BH}$-$\sigma$ relations. They additionally reported an $M_{\rm BH}$-$\sigma$ relation for their combined (core-S\'ersic + S\'ersic) galaxies which is consistent with the relation found by Graham et al.\ (2011). We find that, for all our core-S\'ersic galaxies, the SMBH masses predicted using the Graham \& Scott (2012) $M_{\rm BH}$-$\sigma$ relations which are defined by the core-S\'ersic, S\'ersic or core-S\'ersic + S\'ersic galaxies agree with those used in this study (Table 3) within the range of error bars. We make use of Eq.~4 from Graham et al.\ (2011) and the $\sigma$-values in Table \ref{Tabb1} and assume a 10\% uncertainty on $\sigma$ to estimate the $1\sigma$ uncertainty on each galaxy's SMBH mass (Table \ref{Tabb3}). 

For NGC 4382, Kormendy \& Bender (2009) reported a mass deficit of $M_{\rm
  def}\sim 1.3\times 10^{9} M_{\sun}\approx 13 M_{BH}$ when using
$M_{BH}=1\times10^{8} M_{\sun}$, while G\"ultekin et al.\ (2011) used Nuker
model parameters and found a mass deficit of $M_{\rm def}\sim 5.9\times
10^{8}M_{\sun}\approx 45.6M_{\rm BH}$ when they assumed a very small black hole mass of
$1.3 \times 10^{7} M_{\sun}$.
Ferrarese et al.\ (2006, see their Fig.~20) noted that the large-scale stellar
disk of NGC 4382, which was incorporated into their single S\'ersic fit to
this galaxy, might have biased their $\sim 10 M_{BH}$ mass deficit
measurement. Indeed, their {\it galaxy} S\'ersic index was $\approx 6$--7,
while we have found that the bulge S\'ersic index is less than 3.  Our
analysis of this bulge's central stellar mass deficit yields $M_{\rm
  def}\sim1.2\times10^{8}M_{\sun}\approx1.3M_{\rm BH}$ (using $M_{BH} = 9.55
\times 10^{7} M_{\sun}$), smaller than the results of Ferrarese et al.\
(2006), Kormendy et al.\ (2009) and G\"utekin et al.\ (2011).

Previously, Milosavljevi\'c \& Merritt (2001), Milosavljevi\'c et al.\ (2002),
Ravindranath et al.\ (2002) and Lauer et al.\ (2007a) had used the Nuker model
to measure mass deficits that are up to an order of magnitude larger than
those obtained here.  As detailed in Trujillo et al.\ (2004) and Dullo \&
Graham (2012), this discrepancy arises in part from the differences in the
estimated core sizes from the two models. While the Nuker model break radii are up
to 3 times larger than the core-S\'ersic model break radii, due to the Nuker model fitting a power-law to each galaxy's outer curved S\'ersic profile (Graham et al.\ 2003), the core-S\'ersic break radii agree with model-independent core
sizes where the negative logarithmic slope of the light profile equals 0.5 (Dullo \& Graham 2012).

More recently, Kormendy et al.\ (2009) fit S\'ersic models to core-S\'ersic galaxy light
profiles and tried to quantify the core from a visual inspection, rather than
using the core-S\'esic model in an objective analysis. They report large break
radii and mass deficits $M_{\rm def}\sim 10-20 M_{BH}$.  In an effort to
reconcile these large deficits, they speculated that they may be due to
recoiled black holes, which might enhance the core depletion following their
repetitive core passages (Boylan-Kolchin et al.\ 2004; Gaulandris \& Merritt
2008); cf.\ also Postman et al.\ (2012) for a similar reasoning.  If this
mechanism always occurred, then the lower $M_{\rm def}/ M_{BH}$ ratios in Dullo \& Graham (2012) for elliptical galaxies would suggest that the effective number
of major mergers is less than one, i.e.\ only a minor merger is required.
However given the near-linear relation between black hole mass and host
spheroid mass, and luminosity, for core-S\'ersic galaxies (Graham 2012;
Graham \& Scott 2012) this scenario is unlikely. At least for the elliptical
core-S\'ersic galaxies, which are the dominant population among the known
core-S\'ersic galaxies, this linear black hole-galaxy mass scaling relation readily arises
from the self addition of comparable mass systems, not minor mergers.

\begin{center}
\begin{table*}

\begin {minipage}{135mm}
\caption{Core-S\'ersic galaxy data.}
\label{Tabb3}
\begin{tabular}{@{}llcccccc@{}}
\hline
\hline
Galaxy&$M/L_{V}$&log~($M_{*}/M_{\sun}$)&$R_{e} $&log~($L_{def}/L_{\sun,V}$)&log~($M_{def}/M_{\sun}$)&log~($M_{BH}/M_{\sun}$)\\
&&&(kpc)&&&&\\
(1)&(2)&(3)&(4)&(5)&(6)&(7)\\
\multicolumn{1}{c}{} \\              
\hline                            
NGC 0507  &5.5&11.70&1.65&8.34&9.08&9.22 $\pm$ 0.39\\
NGC 2300   &5.0&11.16&1.02&7.81&8.51&8.86 $\pm$ 0.39\\
NGC 3607  &5.7&11.31&0.84&--&--&--\\
NGC 3706   &5.2&11.60&9.18&8.30&9.01&8.93 $\pm$ 0.39\\
NGC 4382  &2.6&10.90&0.99&7.68&8.09&7.98 $\pm$ 0.38 \\
NGC 6849   &2.4&11.32&2.98&7.97&8.35&8.04 $\pm$ 0.43\\
\hline
\end{tabular} 
Notes.---Col. (1) Galaxy name. Col. (2) $V$-band stellar mass-to-light ($M/L$) ratio.  Col. (3) Stellar mass of the bulge obtained using the bulge magnitude (Table \ref{Tabb1}) and the stellar mass-to-light ($M/L$) ratio (col. 2) (for the elliptical galaxy NGC 3706 the contribution of the galaxy's nuclear stellar ring has not been subtracted). Col. (4) Major-axis half-light radius of the bulge. Col. (5) Central luminosity deficit in terms of $V$-band solar luminosity. Col. (6) Central stellar mass deficit obtained using col. (2) and col. (5). Col. (7) SMBH mass predicted from the $M-\sigma$ relation presented in Graham et al.\ (2011). We use Eq.~4 from Graham et al.\ (2011) and the $\sigma$-values in Table \ref{Tabb1} and a 10\% uncertainty on $\sigma$ to estimate the error on SMBH mass. We adopt a barred morphology to estimate the mass of the black hole in NGC 6849. 
\end {minipage}
\end{table*}
\end{center}

\section{Discussion}\label{Dis}

\subsection {Formation of core-S\'ersic lenticular galaxies}\label{Fr}
As noted above, the central stellar mass deficits for the `core-S\'ersic'
lenticular galaxies are found to be $M_{\rm def}\sim0.5-2 M_{\rm BH}$
(Fig.\ref{Figg4}). By comparing these mass deficits with the results of
Merritt (2006), one would conclude that the bulges of core-S\'ersic S0s have
experienced the equivalent of a few `dry' major merger events. Deviations from
this rule may however exist. For example, substantial black hole recoiling
events would lower this figure of a few, while `loss cone' re-filling through
wet mergers and the production of new stars would allow scope for an increased
number of mergers. This latter scenario is unlikely given, in general, the red
colours of the galaxies' bulges. However, Carter et al.\ (2011) have found
$FUV-NUV$ colour gradients in several core-S\'ersic galaxies which seem to be
inconsistent with the major, dry merger scenario detailed in Faber et al.\
(1997). Also, the globular cluster specific frequencies of some intermediate
luminosity ellipticals are found to be systematically lower than the ones in
bright, core-S\'ersic elliptical galaxies (Harris \& van den Bergh 1981; van den Bergh
1982). This may reflect that core galaxy formation may not involve dry
merger events, because such mergers are thought to be inefficient at creating
new globular clusters.

\begin{figure}
\includegraphics[angle=270,scale=0.55]{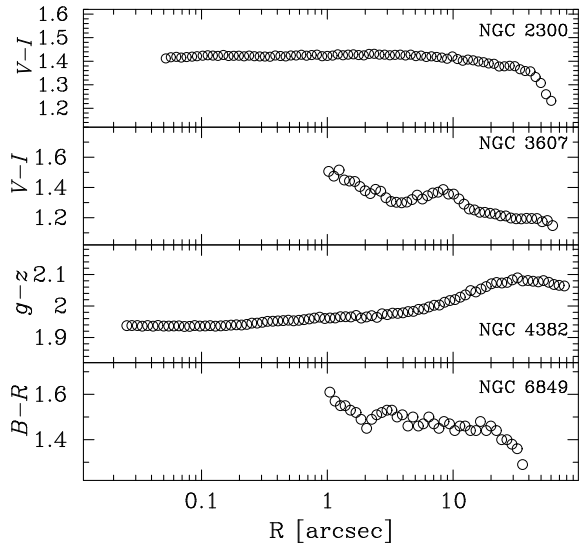}
\caption{Radial colour profiles for four of the five lenticular galaxies
  listed in Table~\ref{Tabb2}. Colour information is not available for
  NGC 507. The inner $R\la 1\arcsec$ profile of NGC 3607 is excluded due to
  possible dust contamination. For NGC 6849, the $B-R$ color profile is
  derived using the published surface brightness profiles presented in Reda et
  al.\ (2004). }
\label{Figg5} 
 \end{figure}

 Although not (yet) popular, various studies have proposed alternative mechanisms
 for generating cores in luminous galaxies. For example, N-body simulations by
 Nipoti et al.\ (2006) revealed that disipationless collapses in preexisting
 dark matter haloes would naturally produce galaxies which resemble ellipticals galaxies with depleted cores. These galaxies may also exist at $ z = 1.5$ to $2$, however, it is unclear how these cores are protected against
 infalling high-density satellites in the absence of a central massive black
 hole. While the bulges of today's S0s do not all possess partially depleted cores, this does not exclude today's S\'ersic bulges from having existed at  $ z = 1.5$ to $2$. Indeed, as shown by MacArthur et al. (2009, and the references in Graham 2013) the bulk of most bulges' stars are sufficiently old to have existed then.
 
 Using numerical simulations, Goerdt et al.\ (2010)
 showed that the energy transferred from sinking massive objects can generate cores that are up to 3 kpc
 in sizes. Similarly, Martizzi et al.\ (2012a, see also Martizzi et al.\ 2012b) used 500 pc resolution 
 simulations and concluded that the combined effects of AGN feedback and
 inspiraling massive black holes create cores in luminous elliptical galaxies
 that are up to 8 - 10 kpc in sizes. However, these overly large cores
 reported by Goerdt et al.\ (2010) and Martizzi et al.\ (2012a) are in general
 incompatible with the majority of $\la 0.5$ kpc cores observed in real
 galaxies (e.g., Trujillo et al.\ 2004; Richings et al.\ 2011; Dullo \& Graham
 2012).

 One can envisage that a core in a core-S\'ersic S0 galaxy may be created
 through the `dry' intermediate-mass merger (with mass ratio 1:4 to 1:7) of an S0 galaxy with a smaller elliptical galaxy, such that the disk is not destroyed but the bulge grows while the black hole binary created during the merger ejects stars from the inner regions of the bulge. Eliche-Moral et al.\ (2012; see also Bekki 1998; Bournaud
 et al.\ 2005) has recently explored this merger scenario and concluded that `dry'
 intermediate-mass mergers can give rise to
 remnants that are both photometrically and kinematically compatible with the
 observed S0s. It should, however, be noted that these studies did not explore the actual nuclear structure of the merger remnants. Another possibility may be minor disk galaxy mergers which result in a rotating, disky merger remnant (Burkert \& Naab 2005; Jesseit et al.\ 2005). Core-S\'ersic S0 galaxies might also or instead be assembled via
 an inside-out mechanism: an early major merger might create an elliptical
 galaxy with a depleted core, around which a disk is subsequently built up
 over time via `dry' minor mergers and/or smooth cold gas accretion. The assumption here is that the accretion process is very gentle and slow, enabling the cold gas to assemble into a star-forming disk (e.g., Steinmentz \& Navarro 2002; Birnboim \& Dekel 2003; Kere\v{s} et al.\ 2005) and also preventing it from falling towards the centers of the bulges, which otherwise would fuel a central burst of star formation and replenish the depleted cores. 

 These later core-S\'ersic galaxy formation scenarios, each involving merger events, may be
 in agreement with the properties of our high luminosity S0 bulges. This is in contrast to
 models that produce S0s by fading the disks of spiral galaxies through
 processes such as ram pressure stripping (e.g., Gunn \& Gott 1972)---which is not to say that such processing does not additionally occur. 
\subsection{Bulges of today's S0s versus  compact high redshift  galaxies}\label{hz}
\begin{figure*}
\includegraphics[angle=270,scale=0.90]{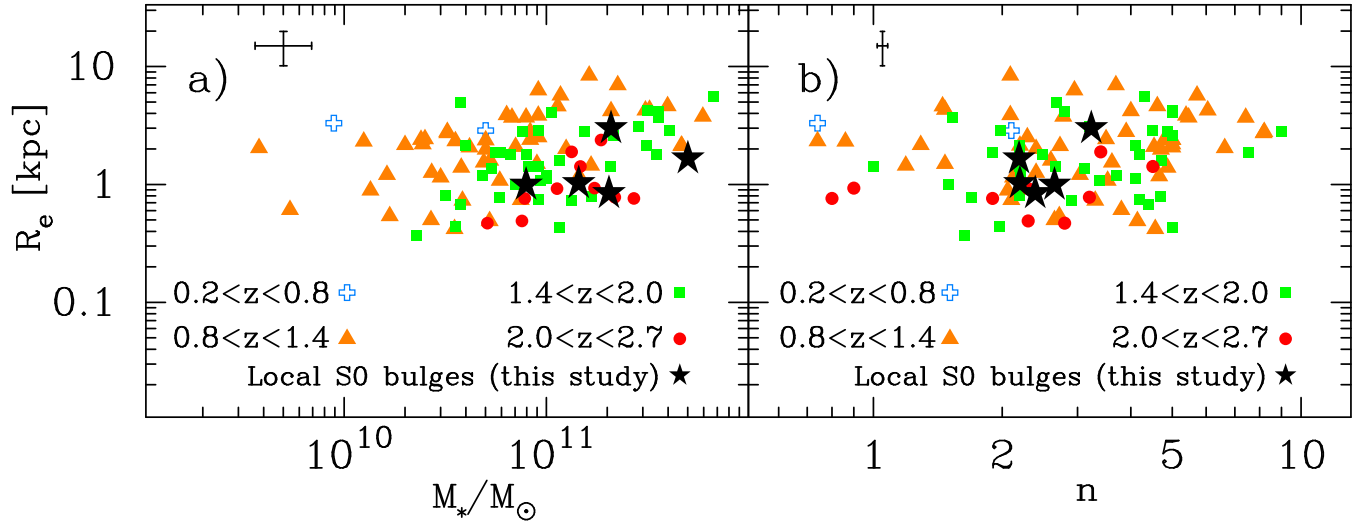}
\caption{Major-axis effective radius $R_{e}$ plotted against (a) stellar mass $M_{*}$,
  and (b) S\'ersic index $n$ for a sample of 106 galaxies. 101 of these
  objects are massive compact galaxies, in the redshift range $0.2 < z < 2.7$,
  published by Damjanov et al.\ (2011, their Table 2 excluding galaxies (i)
  with no reported $n$ values, (ii) fitted with the de Vaucoulers ($R^{1/4}$)
  model, and (iii) modelled by setting $n$ to some constant value). The
  remaining five are local S0 bulges from this study (Table~\ref{Tabb2} and
  \ref{Tabb3}). Median uncertainty from Damjanov et al.\ (2011) for the 101 galaxies is shown by the error bar at the top of each panel.\\\\\\}
\label{Figg6} 
 \end{figure*}

Recent observations have revealed that the sizes of compact massive galaxies
at redshift of $z\sim$ 1.5 to 2 (having effective radii $R_{e}\la$ 2 kpc) are up to a
factor of $\sim 5$ smaller than the present day elliptical galaxies of
comparable stellar mass (Daddi et al.\ 2005; Trujillo et al.\ 2006; Damjanov
et al.\ 2009; Saracco et al.\ 2011; Szomoru et al.\ 2012, and references
therein). While today's elliptical galaxies are widely regarded as the
descendants of these compact high redshift galaxies (e.g., Hopkins et al.\
2009), Graham (2013, his Fig.~1) showed that massive local disk galaxies can
also be alternative candidates given the high stellar density and compactness of their
bulges. Comparing the properties of our massive present day bulges with compact high
redshift galaxies, we plot the size-mass (Fig.~\ref{Figg6}a) and
size-(S\'ersic index) (Fig.~\ref{Figg6}b) diagrams for a sample of 106
galaxies. 101 of these objects are compact massive quiescent galaxies at
redshift $z=$ 0.2 to 2.7 taken from Damjanov et al.\ (2011, selected from
their Table 2), while five are the bulges from local S0s (Table~\ref{Tabb2}
and \ref{Tabb3}). This figure clearly shows that the location of the bulges of
our five core-S\'ersic S0s coincides with that of the high-redshift compact galaxies in both
the $R_e-M_{*}$ and $R_{e}-n$ diagrams.

This overlap
suggests that the compact high redshift galaxies are the
progenitors of (the bulges of) massive present day S0s (e.g., Dutton et al.\ 2012; Graham et al.\ 2013). Indeed, the inside-out growth
mechanism described at the end of Section \ref{Fr}, which could drive the build up of the core-S\'ersic S0s (Section
\ref{6.1.2}), would also eliminate the difficulties encountered in trying to
explain the size evolution of these compact high redshift galaxies into much
larger, modern day elliptical galaxies. Large elliptical galaxies already exist at $z \approx 1.5$ to 2 (e.g., Bruce et al.\ 2012). We further note that van der Wel.\ (2011, see also Chevance
et al.\ 2012) wrote that the majority of compact high redshift galaxies have
small undeveloped disks. In addition, Poggianti et al.\ (2012) found that 4.4\%
of their total low redshift galaxy population had sizes and mass densities
comparable to the compact, massive high redshift galaxies, with 70\% of these
compact low redshift galaxies found to be S0s. Given the results in Graham (2013, his Fig.~1), a comparison of local S0 galaxy ``bulge'' sizes and densities, rather than the entire ``galaxy'' sizes and densities,
 should yield a higher percentage match. It seems plausible that minor mergers and cold gas accretion (e.g.,
Conselice et al.\ 2012), in a preferred plane due to known cosmological
streaming/feeding paths, may transform some of the compact high redshift galaxies into
modern S0s by the creation of a younger surrounding disk. This process eliminates the difficulties that arise when trying to evolve the compact high-$z$ galaxies into local elliptical galaxies, such as the shortage of satellites required to produce the necessary expansion via many minor mergers (Trujillo et al.\ 2012). It is also consistent with cosmological models (e.g., Steinmetz \& Navarro 2002), and explains many properties of early-type galaxies.

\subsection{The role of galaxy environment}\label{6.1.2}
%
Several studies have highlighted that environment plays a major role in the
formation of S0 galaxies (e.g., Dressler 1980; Dressler et al.\ 1997; Couch et
al.\ 1998; Fasano et al.\ 2000; Poggianti et al.\ 2009; Wilman et al.\ 2009;
Just et al.\ 2010; Bekki \& Couch 2011). Three of our five S0 galaxies (NGC
507, NGC 2300 and NGC 3607) reside in X-ray bright galaxy groups, while NGC 6849 is
an isolated galaxy, and NGC 4382 is a member of the Virgo cluster.

 NGC 6849, the isolated core-S\'ersic lenticular galaxy, may have had its
 bulge built through an early violent `dry' major merger, while subsequent
 late accretion of gas and stars built up its disk---the picture in Graham
 (2013), see also Conselice et al.\ (2012).  This hierarchical inside-out
 growth scenario is supported by its radial colour profile plotted in
 Fig.~\ref{Figg5}. The $B-R$ colour gradient reveals that the galaxy becomes
 progressively bluer towards larger radii ($ \ga 10 \arcsec $) where the disk
 dominates. Arnold et al.\ (2011) proposed such a two phase assembly mechanism
 to explain the isolated field galaxy NGC~3115 (except for its dwarf
 companion) based on their analysis of its kinematics and metallicity. Reda
 et al.\ (2004) also found that mergers are a dominant formation path to
 forming some isolated early-type galaxies. In passing we note that van den Bergh
 (2009) alternatively suggested that disk gas ejection by AGN might be a
 dominant process that transforms isolated spirals galaxies into S0s. While
 by itself this process does not appear capable of generating a depleted
 stellar core in galaxies, it may have transformed some spiral galaxies into
 lenticular galaxies.

The `dry' major merger scenario for the bulges of bright lenticular galaxies in
groups 
may be naturally consistent with the observed morphology density relation for
the galaxy groups. Studies have shown that galaxy-galaxy interaction and
merging play an important role in transforming group spirals into S0s and
giving the observed S0 fractions in groups
(e.g., Postman \& Geller1984; Helsdon \& Ponman 2003; Just et al.\ 2010; Bekki
\& Couch 2011). Our three X-ray bright group galaxies are the brightest
central galaxies from their respective groups. As pointed by Helsdon \& Ponman
(2003 and references therein) these central S0 galaxies are assumed to be the
byproducts of an earlier merger activity in the collapsing group core plus
some recent merger and accretion events. Interestingly, as shown in
Fig.~\ref{Figg5}, the outskirts of these galaxies (colour information was not
available for NGC 507) are relatively bluer than their centers. This result
further strengthens the view that bulges in luminous group S0s grow by mergers
while their disks are gradually built and enhanced via latter gas accretion.

For NGC 4382, our fifth lenticular galaxy and the only cluster galaxy in our
sample, our analysis of its brightness profile and thus the mass deficit
(Figs.~\ref{Figg3} and \ref{Figg4}) suggests a `dry' merger event as a
formation path. In contrast, processes such as ram pressure stripping (Gunn \&
Gott 1972; Quilis et al.\ 2000), galaxy harassment (Moore et al.\ 1996) and
strangulation (Larson 1980; Bekki et al.\ 2002) are often noted as plausible
formation mechanisms for cluster lenticular galaxies, rather than galaxy
mergers, due to the high cluster velocity dispersion and high intracluster
medium density. We do not deny these mechanisms, only that they alone cannot
account for partially depleted cores. For example, Vollmer et al.\ (2008a,
2008b) present wonderful observational evidence of the ram pressure stripped Virgo
spiral galaxies NGC 4501 and NGC 4522, while Merluzzi et al. (2012) describe in detail the stripping
process of a galaxy in Abell 3558. The merger event in NGC 4382's past may have occurred prior to it entering the cluster. Coupling N-body simulations with a
semi-analytic formation model, Okamoto \& Nagashima (2001) also reported that
the fractions of cluster S0 galaxies produced by major mergers are
significantly smaller than the observed fractions. In contrast, Bekki (1998)
proposed unequal galaxy merging between two spirals as a dominant formation
origin for cluster lenticular galaxies (see also Burkert \& Naab 2005 and Jesseit et al.\ 2005).
 NGC 4382 is situated in
the outskirts of the Virgo cluster, a location where mergers are likely to
occur, and it displays stellar shells in its image (Schweizer \& Seitzer
1992). These appear to suggest that NGC 4382 might have a merger related
origin (although see Chung et al.\ 2009). The $g-z= 1.95$ bulge colour of NGC
4382 (Fig.~\ref{Figg5}) is however bluer than the typical $g-z= 1.56$ (AB mag)
colour quoted for ellipticals (Fukugita et al.\ 1995). If the core in NGC 4382
was formed from the inspiral of supermassive black holes in a relatively gas-free merger event,
then the progenitor stars which make the bulge were not old, suggestive of
ongoing core formation in some lenticular galaxies rather than all being
formed at redshifts beyond 1.5 to 2 prior to the subsequent accretion of a
disk.

\section{Conclusions} \label{Con2}
We have used the IRAF {\sc ellipse} task to derive the major-axis surface
brightness profiles and isophotal parameters for six early-type galaxies
observed with the high-resolution {\it HST} WFPC2 and ACS cameras.  While one
of these turned out to be an elliptical galaxy, which we modelled with a
core-S\'ersic profile, we have modeled the surface brightness profiles of the
remaining five lenticular galaxies using a core-S\'ersic
model for the bulge plus an exponential model for the disk.  This is the first
time this has been done for disk galaxies with depleted cores.  In our
analysis, we have additionally accounted for the bar component in one of these
disk galaxies by using the Ferrers (1877) function. Our primary conclusions are as follows:\\

1. The core-S\'ersic
bulge plus exponential disk model gives an accurate description to core-S\'ersic lenticular galaxy
 light profiles. The rms residual
scatter of the fits are $\la 0.05$ mag arcsec$^{-2}$.

2. The S\'ersic index $n$ is $\sim3$ for the bulges of our core-S\'ersic
lenticular galaxies, whereas $n\ga3$ for the brighter core-S\'ersic elliptical
galaxies in Dullo \& Graham (2012).

3. The core `break radii' range from 24 to 102 pc.
  
4. We have measured central stellar mass deficits in four of the luminous
``core-S\'ersic'' lenticular galaxies, finding $ M_{\rm def}\sim 0.5-2$
$M_{\rm BH}$, in agreement with previous core-S\'ersic analysis of elliptical
galaxies. (NGC 3607 was excluded because its dusty nuclear spiral compromises
the recovery of its core parameters.)

5. Our results tentatively suggest that, regardless of their environments,
core-S\'ersic lenticular galaxies could be assembled in two stages: an earlier
violent `dry' major merger process involving massive black holes, which forms the bulge component, followed
by subsequent disk formation through minor mergers and/or very gentle cold gas accretion 
about a preferred plane.
Dry intermediate-mass ratio mergers of S0s with smaller BH-hosting ellipticals
could also account for the buildup of some core-S\'ersic S0s.

6. The location of the bulges of our five S0 galaxies, including the S0 NGC 3607 possibly with out a depleted core, in the mass-size and
mass-(S\'ersic index) diagram coincides with that of the compact early-type
galaxies seen at redshift of $z\sim$ 0.2 to 2.7. This could be an indication
that today's massive bulges may be descendants of the compact high redshift
early-type galaxies.

\section{Acknowledgments}
This research was supported under the Australian Research Council’s funding
scheme (DP110103509 and FT110100263). This research has made use of the
NASA/IPAC Extragalactic Database (NED) which is operated by the Jet Propulsion
Laboratory, California Institute of Technology, under contract with the
National Aeronautics and Space Administration.  We are grateful to Juan P. Madrid for his help with IRAF.

\section{Appendix  A} \label{App2}

\subsection{Notes on individual galaxies}\label{Sec42}
In this section we review several relevant features associated with each galaxy in our sample. We
comment on the X-ray properties as this will be relevant to our later discussion about formation scenarios and environment. 
 
\subsubsection{NGC 507}
NGC 507 is the brightest galaxy in the nearby, poor NGC 507 group; it is also
one of the brightest known X-ray early-type galaxies. Laurikainen et al.\ (2010)
found evidence of a weak bar in their $K_{s}$-band image. However, they fitted
a S\'ersic bulge plus an exponential disk to their deep sub-arcsec resolution
$K_{s}$-band data because the bar was too weak to be included in the fit. Our
high-resolution {\em HST}/F555W surface brightness profile for this galaxy is
well fit by the core-S\'ersic bulge + exponential disk model with an rms
residual of 0.027 mag arcsec$^{-2}$ (Fig.~\ref{Figg3}).

\subsubsection{NGC 2300}
NGC 2300 belongs to the poor NGC 2300 group. Mulchaey et al.\ (1993) reported
the presence of hot diffuse intragroup gas close to NGC 2300 using {\em
  ROSAT}.  Although several previous studies (e.g., Huchtmeier 1994, Sandage
\& Bedke 1994) have classified this galaxy as an elliptical, as shown in
Fig.~\ref{Figg3}, the luminosity profile is well fit by the core-S\'ersic
bulge + exponential disk model (see also Laurikainen et al.\ 2010 for their
S\'ersic + exponential fit to ground-based data that does not resolve the
depleted core.)

\subsubsection{NGC 3607}
NGC 3607 is an X-ray luminous (e.g., Terashima et al.\ 2000), bright
lenticular galaxy (e.g., de Vaucouleurs et al. 1991; Afanasiev \& Silchenko
2007; Laurikainen et al.\ 2010; Guti\'errez et al.\ 2011; although it is
classified as an elliptical in G$\ddot{\mbox{u}}$ltekin et al.\ 2009) located in the
Leo II group. It has a dust obscured central region as can be seen from the
residual image in Appendix B, Fig.~\ref{Figg7}. We constructed an image model using
the IRAF task {\sc bmodel} and subtracted it from the raw {\it V}-band image
to obtain this residual image. The effect of this dusty nuclear spiral can also be seen in the position angle, ellipticity and isophote shape parameter ($B_{4}$) profiles (Fig.~\ref{Figg2}). In addition to using this to guide the careful
dust masking procedure, performed prior to running {\sc ellipse} to obtain the $V$-band light profile
 beyond 1$\arcsec$, we have analysed the galaxy's near-infrared {\em HST}/NICMOS
F160W ({\it H}-band) image which is less affected by dust. As noted in Section
2, the inner ($R \la 1\arcsec$) {\it V}-band profile of this galaxy, taken
from Lauer et al.\ (2005), is derived from a PSF-deconvolved image. Therefore,
the apparent 1$\arcsec$.31 core of the {\it V}-band profile fit
is most likely an artifact due to interference from the dusty nuclear spiral
disk with the deconvolution routine (as noted already by Dullo \& Graham 2012,
and references therein).

\subsubsection{NGC 3706}
In agreement with the ground-based work by Laurikainen et al.\ (2010), the
analysis of NGC~3706's {\em HST}/F555W light profile suggests that it is
likely an elliptical galaxy rather than a lenticular galaxy. The
unsharp-masked image (Appendix B, Fig.~\ref{Figg7}), the position angle twist, the disky and very flat (`pointy') nature of the isophotes from the {\sc ellipse} fit in the region
$R$= $0\arcsec .2-1\arcsec$.0 (Fig.~\ref{Figg2}) reveal the presence of an
edge-on nuclear ring of stars in this galaxy (Lauer et al.\ 2002; Kandrup et
al.\ 2003). We model the host galaxy light profile after excluding the region
affected by this additional ring of star light (Fig.~\ref{Figg3}).

\subsubsection{NGC 4382}
NGC 4382 is a peculiar, fast rotating galaxy (Emsellem et al.\ 2007) in the
Virgo cluster which is classified as an S0 in the RC3. Based on the photometry
and brightness profile analysis, but not the rotational evidence, Kormendy et
al.\ (2009) reported an absence of a large-scale stellar disk in this galaxy,
and identified the extra light at large radii to be a feature associated with
a recent merger (although see Chung et al.\ 2009), hence, they classified the
galaxy as E2. In contrast, Laurikainen et al.\ (2011) have classified the
galaxy as an S0 based on the detection of dispersed spiral arm segments,
although they remained uncertain about the exponential nature of the disk light
distribution. In Fig.~\ref{Figg3}, we show that the galaxy light profile is
indeed well fit by the core-S\'ersic bulge + exponential disk model with an rms
residual of only 0.016 mag arcsec$^{-2}$.  This galaxy has positive $B_{4}$
values, i.e.\ disky isophotes where $R > 1\arcsec$. We also note that we
measure the galaxy's bulge-to-disk (B/D) flux ratio to be $\sim$0.67
(B/T=0.40), revealing the existence of an appreciable fraction of the galaxy
light in the large-scale disk. Dullo \& Graham (2012, their Fig.~7) show how
this large-scale disk  results in a different ``galaxy'' S\'ersic index, from the single-component fit in
Ferrarese et al.\ (2006), compared to the ``bulge'' S\'ersic index. Not fitting for the outer disk results in a `galaxy'
S\'ersic index which is larger than the ``bulge'' S\'ersic index and thus
over-predicts the central mass deficit.

\subsubsection{NGC 6849}\label{Sec416}
NGC 6849 is classified as an isolated, barred lenticular (SB0) galaxy in the
RC3.  It is also classified as an elliptical galaxy in the ground-based
studies of Saraiva et al.\ (1999) and de Souza, Gadotti \& dos Anjos (2004).
In contrast to the findings of Saraiva et al.\ (1999), the F555W/{\em HST}
unsharp-masking (Appendix B, Fig.\ \ref{Figg7}) appears to suggest the presence
 of a bar in this galaxy.
Fig.~\ref{Figg2} illustrates that the ellipticity of this galaxy increases
towards larger radii, which makes it the flattest in our sample.  As shown in
Fig.~\ref{Figg3}, the 3-component bulge-bar-disk luminosity profile is well
fit by the the core-S\'ersic bulge plus Ferrers bar plus exponential disk
decomposition model with an rms residual of $\sim $0.04 mag arcsec$^{-2}$.

\subsection{Comments on Dullo \& Graham (2012)} \label{Sec422}
\begin{center}  
\begin{figure*}
\begin {minipage}{178mm}
~~~~~~~~~~~\includegraphics[angle=270,scale=0.60]{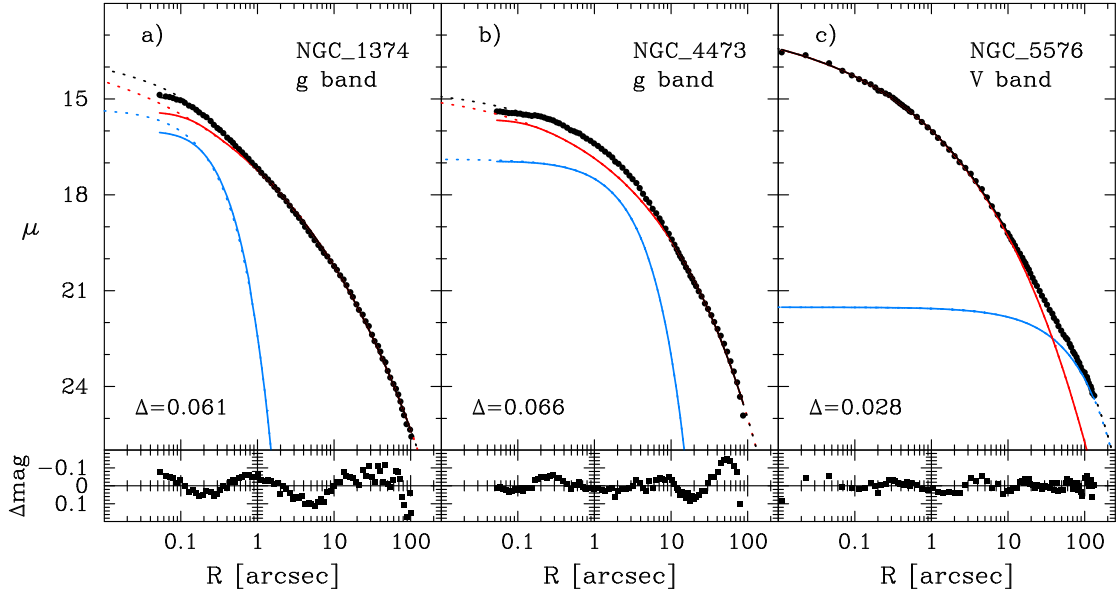}
\caption{Double S\'ersic model fits to the surface brightness profile of NGC 1374 (inner exponential + S\'ersic), NGC 4473 (inner exponential + S\'ersic) and NGC 5576 (inner S\'ersic + outer S\'ersic), see the text for details. The $\sim 100 \arcsec$ geometric mean light profiles for NGC 1374 and NGC 4473 were taken from Turner et al.\ (2012) and Ferrarese et al.\ (2006), respectively. For NGC 5576, we use the Lauer (2012) composite ( deconvolved {\it HST} data of Lauer et al.\ 2005 at $R \la 10 \arcsec$ + ground-based data at $R \ga 10 \arcsec$) major-axis light profile which extends out to $\sim 100 \arcsec$. In contrast to Lauer (2012, his Fig.~1) and Lauer et al.\ (2005), our modelling (as in Dullo \& Graham 2012) reveals the absence of depleted cores in these galaxies. The rms residuals, $\Delta$, are shown in each panel.}
\label{Figg3L}
\end{minipage}
\end{figure*}
\end{center}

Commenting on Dullo \& Graham (2012), Lauer (2012) presented S\'ersic
model fits to the outer regions of composite light profiles (using the inner
10$\arcsec$ {\it HST} profiles from Lauer et al.\ 2005 combined with
ground-based profiles out to $\sim$100$\arcsec$ taken from the literature) for three galaxies
which Dullo \& Graham (2012), and others,  had shown not to possess depleted cores:
NGC 1374, NGC 4473 and NGC 5576. Lauer's (2012) fits suggested the presence of depleted cores relative to his
outer S\'ersic model. He therefore claimed that Dullo \&
Graham (2012) had misclassified these three galaxies as coreless
S\'ersic galaxies due to an incorrect application of the S\'ersic
model to radially limited 10$\arcsec$ light profiles taken from
Lauer et al.\ (2005). However, there were already $\sim$100$\arcsec$
high-resolution {\it HST} profiles published for these galaxies that
had been well fit using S\'ersic models with no central light deficit
(Turner et al.\ 2012, NGC 1374; Ferrearese et al.\ 2006 and Kormendy
et al.\ 2009, NGC 4473; Trujillo et al.\ 2004, NGC 5576).
Fig.~\ref{Figg3L} shows that the light profiles for these galaxies,
covering a large radial extent, {\it are} described by the sum of two
S\'ersic profiles without any partially depleted core.  The reason for
these two-component models is detailed below. Basically, taking into
account additional information, such as the presence of tidal material and/or
kinematic substructure can help to identify the required components of a
fit. In general, we recommend showing the residuals about one's fitted model, as 
they can reveal if the large-scale curvature in the radial light distribution is well 
matched, and we advise against subjectively restricting the fitted radial range.

\subsubsection{NGC 1374}
Turner et al. (2012) fit a PSF-convolved S\'ersic model to this galaxy, with no evidence for a depleted core. In Dullo \& Graham (2012) we revealed that there is actually an additional component at the center of this galaxy. In Fig.~\ref{Figg3L}a, we show our (PSF convolved) nuclear disk + S\'ersic model
fit to the light profile of NGC 1374 taken from Turner et al.\ (2012), see
also the fit presented in Dullo \& Graham (2012) to the inner $\sim 10 \arcsec$. Some small (in amplitude), large-scale residual is present in Fig.~\ref{Figg3L}a. The rotation curve in Graham
et al.\ (1998, see also Longo et al.\ 1994) revealed that this galaxy rotates out
to $40\arcsec$, suggesting the presence of an additional component which may be the cause of some of the residual structure about our exponential (for the nuclear disk) plus $n=4.3$ S\'ersic (for the underlying host galaxy light) model. 

\subsubsection{NGC 4473}
Ferrarese et al.\ (2006) fit a PSF-convolved S\'ersic model to NGC 4473, with no depleted core. Kormendy et al.\ (2009) 
identified a central light excess in this galaxy, above their
adopted S\'ersic ($n\approx 4$) fit, and they remarked that the extra light is due
to a known counter-rotating embedded stellar disk (Emsellem et al.\ 2004; Cappellari \& McDermid 2005; Cappellari et al.\ 2007).  Fig.~\ref{Figg3L}b  shows our convolved exponential disk + S\'ersic bulge model for this galaxy's light profile taken from Ferrarese et al.\ (2006). We fit an exponential $n=1$ function to the counter rotating disk, and find that a S\'ersic $n\sim3$ model fits the underlying galaxy light. Furthermore, Pinkney et al.\ (2003) remarked that this galaxy has unusual parameters from their Nuker model fit. They found that it has i) the smallest $\alpha$ value of their sample (i.e. a broad transition region between the inner and outer Nuker model power-law profiles) and ii) a very steep $\beta$ value (the steepest from all the galaxies modelled by Byun
et al.\ 1996). As warned in Graham et al.\ (2003), this is what one expects when fitting the Nuker model to what is actually  a S\'ersic profile with a low value of $n$ and no depleted core. Pinkney et al.\ (2003) additionally remarked that the absolute magnitude of this galaxy is consistent with the ``power-law'' galaxies, i.e. those without depleted cores.
\subsubsection{NGC 5576}
Trujillo et al.\ (2004) modelled the deconvolved 100$\arcsec$ light profile of NGC 5576 with the S\'ersic model, finding no depleted core. Kim et al.\ (2012, their Fig.~5) showed that NGC 5576 and the barred S0
NGC 5574 are an interacting pair (see also Tal et al.\ 2009) having a long
tidal tail, which extends out to $60$ kpc in their 3.6 $\mu $m image. They
 showed the presence of tidal disturbances in the outskirts (from
1$\arcmin$ to $2\arcmin$) of NGC 5576 because of this interaction.
In Fig.~\ref{Figg3L}c, we therefore fit a double S\'ersic model to the  light profile of NGC 5576 presented in Lauer (2012)\footnote{Lauer (2012) noted that his ground-based profile for NGC 5576 was provided by Michard \& Marcheal (1993), however this galaxy was not actually included by those authors.}. We find that the central galaxy light distribution is well fit with an inner $n \sim 3.5$ S\'ersic model, and the extra light, which is likely to be due to  the re-distribution of material from the minor to the major-axis as a result of the tidal interaction, is described by an outer $n \sim 1.2$ S\'ersic model. Kim et al.\ (2012) also fit a double S\'ersic model (an inner $n=3.45$ S\'ersic model plus an outer $n=1.48$ S\'ersic model) to their  light profile which is sampled from $R= 11\arcsec.8$ to $247\arcsec.5$.  In general agreement with this and the inner $n \sim 4.5$ S\'ersic model fit presented in Trujillo et al.\ (2004), our $n \sim 3.5$ S\'ersic fit to this galaxy's inner light distribution (Fig.~\ref{Figg3L}c) not only shows the absence of a central luminosity deficit in this galaxy, but is also consistent with its magnitude $M_{V}= -21.11$ mag and velocity dispersion $\sigma=171$ km s$^{-1}$ (see Graham et al.\ 2001, their Fig.~13; Dullo \& Graham 2012, their Fig.~14). This is in contrast to the single-component $n \sim 8$ S\'ersic fit by Lauer (2012, their Fig~1).

We also note that, contrary to the claim by Lauer (2012), Dullo \&
Graham (2012, their section 4) explicitly noted that the spurious
downward departure in the inner light profile of NGC 4486B (relative
to the inward extrapolation of its outer S\'ersic profile) is due to
the presence of a double optical nucleus, rather than a typical
depleted core. Coupled with NGC 4458, NGC 4478 and NGC 7213, we
therefore have that seven of the 39 galaxies (i.e.\ $\sim$18\%)
classified as ``core'' galaxies according to the Nuker model analysis
by Lauer et al.\ (2005), do not have central stellar deficits relative
to their outer S\'ersic profiles and are therefore not identified as
``core-S\'ersic'' galaxies. Such `mismatches' will be most prevalent
among galaxy samples fainter than $M_{B} \approx -20.5$ mag that
contain spheroids with low S\'ersic indices and thus have relatively
flat inner cores that are not depleted of stars.

The ATLAS 3D galaxy sample used by Lauer (2012) is dominated by
early-type galaxies with magnitudes up to $\sim$3 mag fainter than
$M_{B} = -21$ mag --- many of which are two-component lenticular
galaxies (Emsellem et al.\ 2011).  As such, application of the Nuker
model may identify many `cores' ($\gamma < 0.3$) among these galaxies
which do not actually possess partially-depleted cores relative to
their spheroid's outer S\'ersic profile.  The Nuker model
classification may therefore substantially blur the actual connection
between the dry merger scenario, as described by Faber et al.\ (1997),
and the existence of depleted cores from coalescing black holes.  We
thus advocate the core-S\'ersic model and point readers to additional
important reasons for this that are discussed in Graham et al.\ (2003)
and Dullo \& Graham (2012). 

\section{Appendix  B}
\begin{figure}
~~~~\includegraphics[angle=0,scale=0.265]{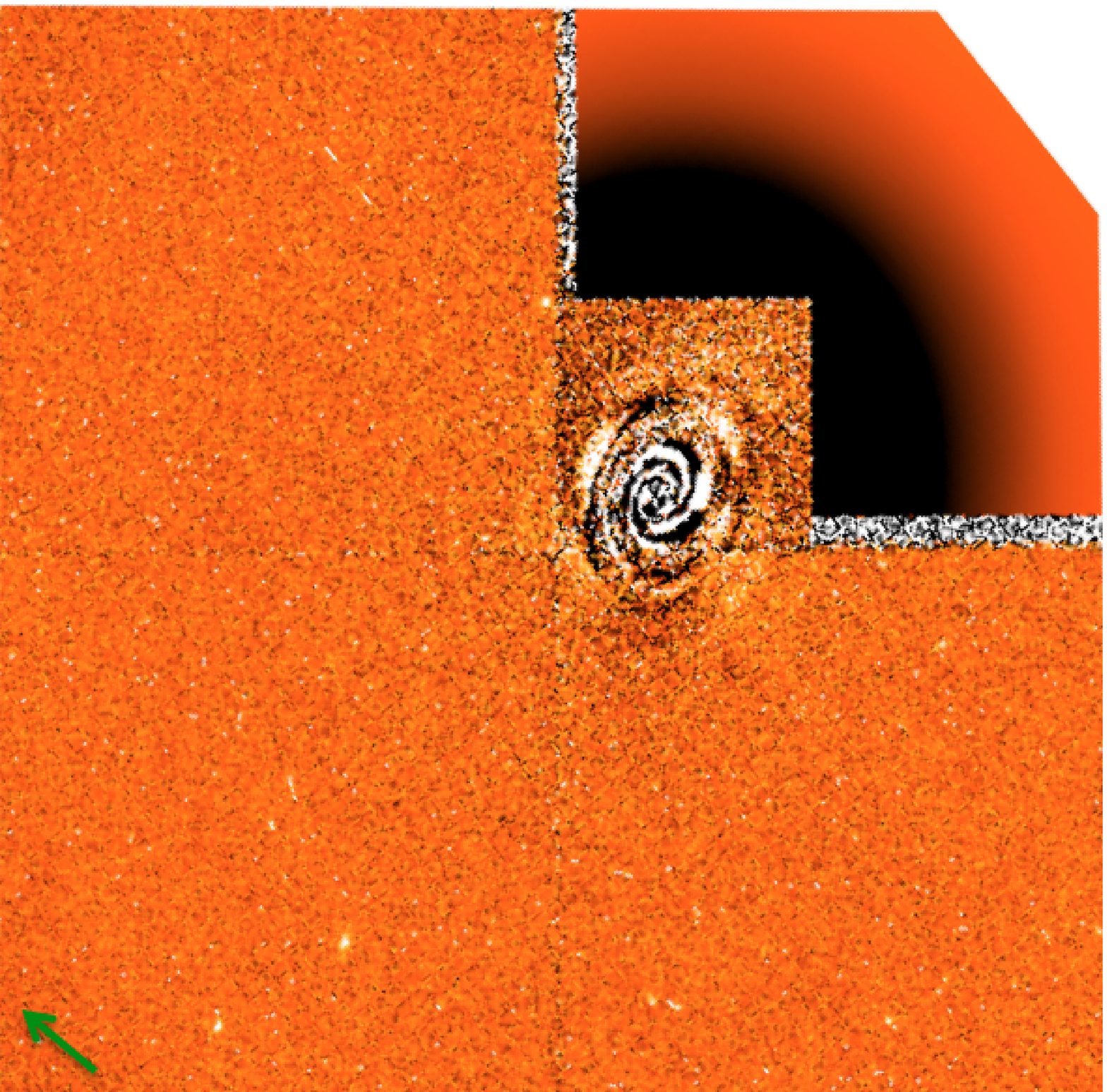}
\includegraphics[angle=0,scale=0.1620]{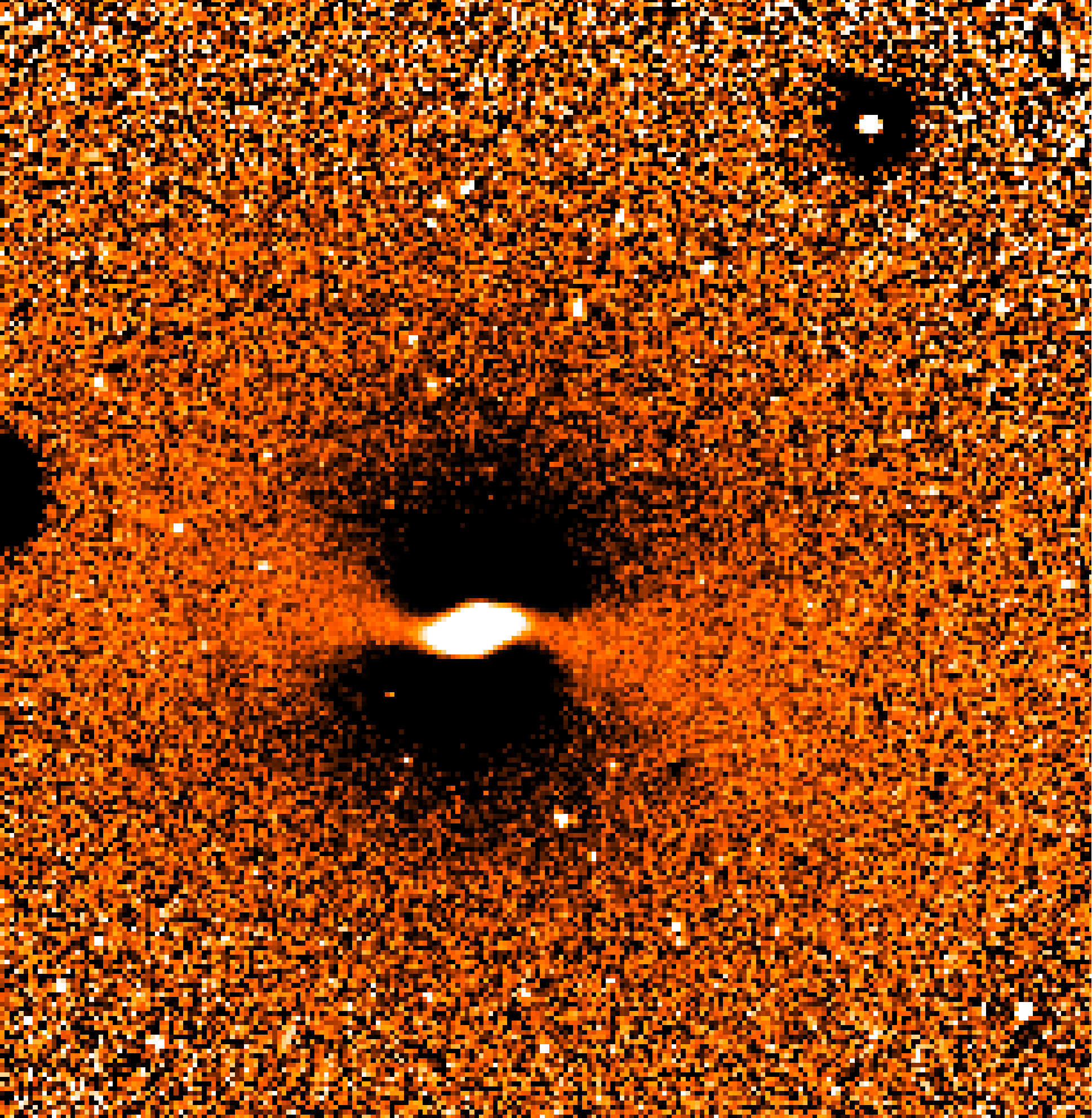}
\includegraphics[angle=0,scale=0.54]{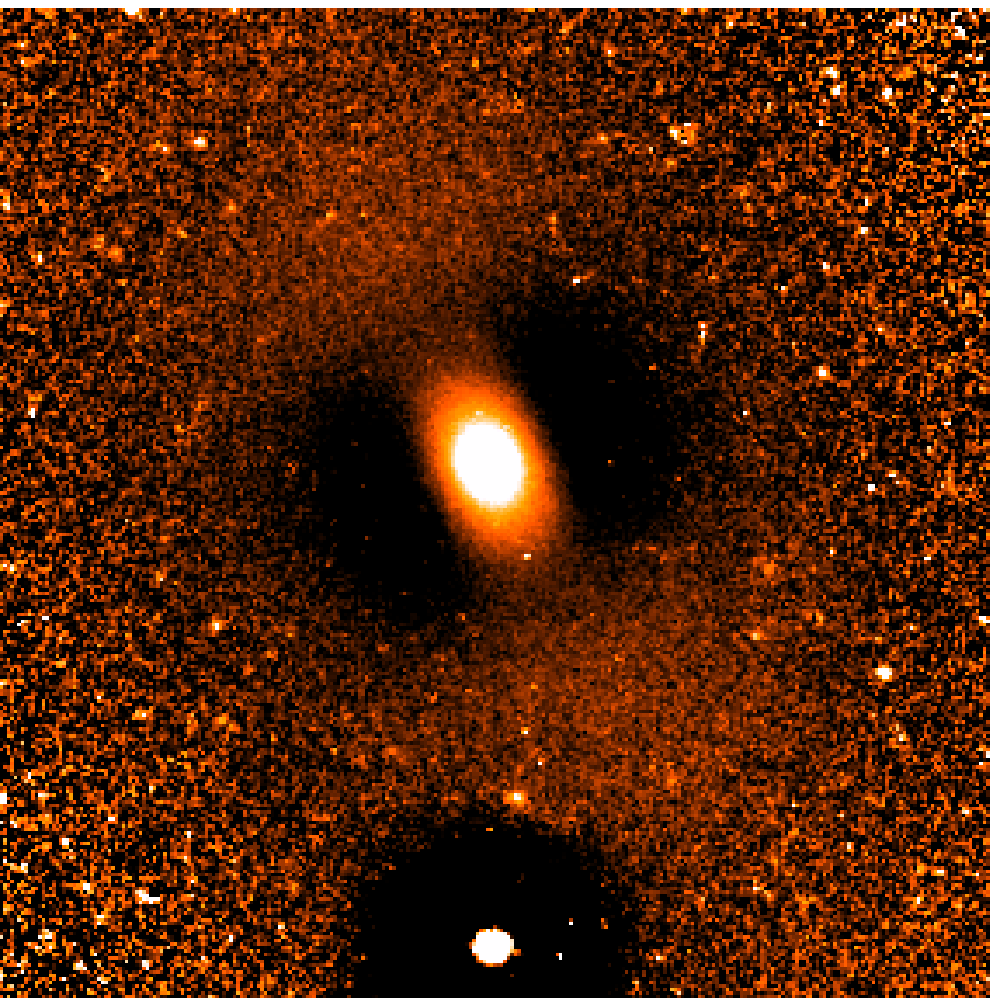}

\caption{Left: Residual image of NGC 3607 obtained by subtracting our symmetrical image model from the original {\em V}-band image of the galaxy observed using the {\em HST}/WFPC2 camera. A dust disk is seen in the central region of the galaxy, see the Appendix A text for a further description. Middle: {\it HST} WFPC2/F555W PC 15$\arcsec$$\times$$15\arcsec$ unsharp masked image of NGC 3706 showing an edge-on nuclear stellar ring (see Fig.~\ref{Figg3} and Lauer et al.\ 2002, their Figure 1). Right: {\it HST} WFPC2/F555W PC 15$\arcsec$$\times$15$\arcsec$ unsharp masked image of NGC 6849 showing an inner bar. For NGC 3607, north is in the direction of the arrow. North is up for NGC 3706 and NGC 6849.}   
\label{Figg7} 
\end{figure}

Fig.~\ref{Figg7} shows distinct features in NGC 3607 (left), NGC 3706 (middle) and NGC 6849 (right).

\label{lastpage}

\begin{references}{}
\reference{afan}Afanasiev, V. L., \& Sil'chenko, O. K. 2007, Astron. Astrophys. Trans., 26, 311
\reference{Andre} Andredakis, Y. C., Peletier, R. F., \& Balcells, M. 1995, MNRAS, 275, 874
\reference{Arnol}Arnold, J. A., Romanowsky, A. J., Brodie, J. P., et al. 2011, ApJ, 736, L26
\reference{barn} Barnes, J. E., Hernquist, L., 1992, ARA\&A, 30, 705
\reference{begl} Begelman, M. C., Blandford, R. D., \& Rees, M. J.\
1980, Nature, 287, 307
\reference{Bell4} Bell, E.F., Wolf C., Meisenheimer, K., et al.\ 2004, ApJ, 608, 752 
\reference{bell06} Bell, E. F. et al., 2006, ApJ, 640, 241
\reference{bekk} Bekki, K. 1998, ApJL, 502, 133
\reference{bekk2} Bekki, K., Couch, W. J.,\& Shioya, Y. 2002, ApJ, 577, 651
\reference{bekki} Bekki, K., \& Couch, W. J. 2011, MNRAS, 415, 1783
\reference{bekki} Bekki, K. \& Graham, A. W. 2010, ApJ, 714, L313
\reference{Ben88} Bender, R., Doebereiner, S., Moellenhoff C. 1988, A\&AS, 74, 385
\reference{Bez} Bezanson, R., van Dokkum, P., \& Franx M. 2012, ApJ, 760, 62
\reference{black} Blakeslee, J. P., Lucey, J. R., Tonry, J. L., Hudson, M. J., Narayanan, V. K.,
Harris, B. J., 2002, MNRAS, 330, 443
\reference{Birb}Birnboim, Y., Dekel, A. 2003, MNRAS, 345, 349
\reference{bourn} Bournaud, F., Jog, C. J., \& Combes, F. 2005, A\&A, 437,
69
\reference{boyl1} Boylan-Kolchin, M., Ma C.-P., 2007, MNRAS, 374, 1227
\reference{boyl2} Boylan-Kolchin, M., Ma C., Quataert, E., 2004, ApJ, 613, L37
\reference{Bluck} Bluck, A. F. L., Conselice, C. J., Buitrago, F., Gr\"utzbauch, R., Hoyos C., Mortlock, A., Bauer, A. E., 2012, ApJ, 747, 34
\reference{Bru} Bruce, V. A., Dunlop, J. S., Cirasuolo, M., et al. 2012, MNRAS, 427, 1666
\reference{burkert} Burkert, A., \& Naab, T. 2005, MNRAS, 363, 597
\reference{byun} Byun, Y.-I., et al. 1996, AJ, 111, 1889
\reference{caon} Caon N., Capaccioli M., D'Onofrio, M.\ 1993, MNRAS, 265, 1013
\reference{capp} Cappellari, M. et al., 2007, MNRAS, 379, 418
\reference{capp} Cappellari, M., McDermid, R. M., 2005, Class. Quantum Grav., 22, 347
\reference{Car78} Carter, D. 1978, MNRAS, 182, 797
\reference{Car87} Carter, D. 1987, ApJ, 312, 514
\reference{Cart} Carter, D., Pass S., Kennedy J., Karick A.M., Smith R.J. 2011, MNRAS, 414, 341
\reference{catt} Cattaneo, A. et al., 2009, Nat, 460, 213
\reference{Cheva} Chevance, M., Weijmans, A., Damjanov, I. et al., 2012, ApJ, 754, 24
\reference{chung} Chung A., van Gorkom J. H., Kenney J. D. P., Crowl H., Vollmer B., 2009,
AJ, 138, 1741
\reference{Cons} Conselice, C. J., et al, arXiv:1206.6995 
\reference{cou} Couch, W. J., Barger, A. J., Smail, I., Ellis, R. S., Sharples, R. M., 1998, ApJ, 430, 121
\reference{cox} Cox, T. J., Jonsson, P., Primack, J. R., \& Somerville, R. S. 2006, MNRAS,
373, 1013
\reference{Chung} Chung, A., van Gorkom, J. H., Kenney, J. D. P., Crowl, H., \& Vollmer, B.
2009, AJ, 138, 1741
\reference{daddi} Daddi, E., Renzini, A., Pirzkal, N., et al. 2005, ApJ, 626, 680
\reference{damj1} Damjanov, I., Abraham, R. G., \& Glazebrook, K. et al. 2011, ApJ, 739, L44 
\reference{damj2} Damjanov, I., McCarthy, P. J., \& Abraham, R. G. et al. 2009, ApJ, 695, 101
\reference{Davie} Davies, R.L., Efstathiou, G., Fall, S.M., Illing- worth, G., Schechter, P.L. 1983, ApJ, 266, 41
\reference{desi}Desai, V., et al., 2007, ApJ, 660, 1151
\reference{de}de Souza, R. E., Gadotti, D. A., dos Anjos, S., 2004, ApJS, 153,
411
\reference{dev48} de Vaucouleurs, G., 1948, Ann. d'Astrophys., 11, 247
\reference{deV91}de Vaucouleurs, G., de Vaucouleurs A., Corwin H. G., Jr., et al.
1991, Third Reference Catalogue of Bright Galaxies (Berlin: Springer)
\reference{dhar1} Dhar, B. K., \& Williams L. L. R., arXiv:1112.3120 
\reference{dres} Dressler, A., 1980, ApJ, 236, 351
\reference{dres2} Dressler, A., Oemler A., Couch W. J., Smail I., Ellis R. S., Barger A., Butcher H., Poggianti  B. M., Sharples R. M., 1997, ApJ, 490, 577
\reference{drive} Driver, S. P., Popescu, C. C., Tuffs, R. J., Graham, A. W., Liske, J., Baldry, I., 2008, ApJ, 678, L101
\reference{DullonGraham}Dullo, B. T., \& Graham, A. W., ApJ, 755, 163 
\reference{dut} Dutton, A. A., Treu, T., Brewer, B. J., et al. 2012, MNRAS, submitted, arXiv:1206.4310
\reference{Ebi} Ebisuzaki, T., Makino, J., \& Okumura, S. K. 1991, Nature, 354, 212
\reference{Elich} Eliche-Moral, M. C., Gonzale-Garcia, A. C., Aguerri, J. A., et al., arXiv:1209.0782
\reference{Emse1} Emsellem, E., Cappellari, M., Krajnovi\'c, D., et al. 2007, MNRAS, 379, 401
\reference{Emse2} Emsellem, E., Cappellari, M., Krajnovi\'c, D., et al.\ 2011, MNRAS, 414, 888  
\reference{Emse3} Emsellem, E., Cappellari, M., Peletier, R. F., et al. 2004, MNRAS, 352, 721
\reference{faber}Faber, S. M., et al.\ 1997, AJ, 114, 1771
\reference{Fasano} Fasano, G., Poggianti, B. M., Couch, W. J., Bettoni, D., Kj\ae gaard, P., Moles, M., 2000, ApJ, 542, 673
\reference{ferr06} Ferrarese, L., et al.\ 2006, ApJS, 164, 334
\reference{ferrndford} Ferrarese, L., Ford H., 2005, Space Sci. Rev., 116, 523
\reference{Fer1994} Ferrarese, L., van den Bosch, F. C., Ford, H. C., Jaffe, W., \& O’Connell, R. W.
1994, AJ, 108, 1598
\reference{F1877}Ferrers, N.M. 1877, Quart.\ J.\ Pure Appl.\ Math., 14, 1
\reference{Forbe}Forbes, D.A., Spitler, L.R., Strader, J., et al.\ 2011, MNRAS, 413, 2943 
\reference{ford} Ford, H. C., Bartko, F., Bely, P. Y., et al. 1998, Proc. SPIE, 3356, 234
\reference{goer}Goerdt, T., Moore, B., Read, J. I., Stadel, J., 2010, ApJ, 725, 1707
\reference{Gove}Governato, F., Brook, C. B., Brooks, A. M., Mayer, L., Willman,
B., Jonsson P., Stilp, A. M., Pope, L., Christensen, C., Wadsley, J.,
Quinn T., 2009, MNRAS, 398, 312
 \reference{Graham01} Graham, A. W., 2001, AJ, 121, 820
 \reference{gra04} Graham, A. W., 2004, ApJ, 613, L33
 \reference{gra07} Graham, A. W., 2007, MNRAS, 379, 711
\reference{Grah2012} Graham, A. W., 2012, ApJ, 746, 113
 \reference{gr11b} Graham, A. W., 2013, to appear in 'Planets, Stars and Stellar Systems'. Springer, Berlin , preprint (arXiv:1108.0997)
\reference{graham08} Graham, A. W., Colless, M. M., Busarello, G., Zaggia, S., \& Longo, G.
1998, A\&AS, 133, 325
 \reference{gra05} Graham, A. W., \& Driver S.P.\ 2005, Publ. Astron. Soc. Australia, 22, 118
\reference{gra03} Graham, A. W., Erwin P., Trujillo I., \& Asensio Ramos A.\ 2003, AJ, 125, 2951
 \reference{GaG03} Graham, A. W., \& Guzm\'an, R.\ 2003, AJ, 125, 2936
 \reference{Graet} Graham, A. W., Onken, C.A., Athanassoula, E., \& Combes, F.\ 2011, MNRAS, 412, 2211
\reference{Grascot13} Graham, A. W., \& Scott, N. 2013, ApJ, 764, 151
\reference{Grahamnspit}Graham, A. W., Spitler, L. R., 2009, MNRAS, 397, 2148
\reference{gratru2001} Graham, A. W., Trujillo, I., Caon, N., 2001, AJ, 122, 1707
\reference{GaW08} Graham, A. W., \& Worley, C.C., 2008, MNRAS, 388, 1708 
\reference{Gualnmerri} Gualandris, A., Merritt, D., 2008, ApJ, 678, 780
\reference{Gualars} Gualandris, A., \& Merritt, D. 2012, ApJ, 744, 74
\reference{Gul09}G{\"u}ltekin, K., Richstone, D.~O., Gebhardt, K., et al.\ 2009, ApJ, 695, 1577
\reference{Gul11}G{\"u}ltekin K., Richstone D.~O., Gebhardt K., et al.\ 2011, ApJ, 741, 38
 \reference{Gunn} Gunn, J. E., Gott, J. R. III., 1972, ApJ, 176, 1
\reference{gutier}Guti\'errez, L., Erwin, P., Aladro, R., \& Beckman, J. E. 2011, AJ,
142, 145
\reference{Harris}Harris, W.E., van den Bergh, S. 1981, AJ, 86, 1627
\reference{helsdon}Helsdon, S. F., \& Ponman, T. J. 2003, MNRAS, 339, L29
\reference{holtz}Holtzman, J. A., Burrows, C. J., Casertano, S., Hester, J. J., Trauger,
J. T., Watson, A. M., Worthey, G., 1995, PASP, 107, 1065
\reference{Hopk1} Hopkins, P. F., Bundy, K., Murray, N., Quataert, E., Lauer, T. R., Ma C.-P.,
2009, MNRAS, 398, 898
\reference{hopk2} Hopkins, P. F., \& Hernquist, L. 2010, MNRAS, 407, 447
\reference{Hub} Hubble, E. 1929, Proceedings of the National Academy of Science, 15, 168
\reference{hubble} Hubble, E.P. 1936, The Realm of the Nebulae, by E.P. Hubble, New Haven: Yale University Press
\reference{Huch}Huchtmeier, W. K., 1994, A\&A, 286, 389
\reference{Hyde}Hyde, J. B., Bernardi M., Fritz A., Sheth R. K., Nichol R. C.,
2008, MNRAS, 391, 1559
\reference{jaffe} Jaffe, W., Ford H. C., O'Connell R. W., van den Bosch F. C., \& Ferrarese L.\ 1994, AJ, 108, 1567
\reference{Jeans} Jeans, J.H. 1928, Astronomy \& Cosmogony, (Cam- bridge: Cambridge University Press), p.332
\reference{Jed}Jedrzejewski, R. I., 1987, MNRAS, 226, 747
\reference{Jess} Jesseit, R., Naab, T., \& Burkert, A. 2005, MNRAS, 360, 1185
\reference{just}Just, D. W., Zaritsky D., Sand D. J., Desai V., \& Rudnick G. 2010, ApJ, 711, 192
\reference{kandrup} Kandrup, H. E., Sideris I. V., Terzi\'c B., Bohn C. L., 2003, ApJ, 597, 111
\reference{Kauuf} Kauffmann, G., \& Haehnelt, M. 2000, MNRAS, 311, 576
\reference{kawa}Kawata, D. \& Mulchaey J. S. 2008, ApJl, 672, L103
\reference{Kers}Kere\v{s}, D., Katz, N., Weinberg, D. H., \& Dav\'e, R. 2005, MNRAS, 363, 2
\reference{Khoch} Khochfar, S., Burkert, A., 2001, ApJ, 561, 517
\reference{kim} Kim, T., Sheth, K., Hinz, J. L., et al. 2012, ApJ, 753, 43 
\reference{korm2} Kormendy, J., \& Bender, R. 2009, ApJ, 691, L142
\reference{Kor09}Kormendy, J., Fisher, D.B., Cornell, M.E., Bender, R.\ 2009, ApJS, 182, 216
\reference{Kraj} Krajnovi\'c, D., Cappellari, M., de Zeeuw, P. T., \& Copin, Y. 2006, MNRAS,
366, 787
\reference{Kulkarni} Kulkarni, G. \& Loeb, A. 2012, MNRAS, 422, 1306
\reference{lars}Larson, R. B., Tinsley, B. M., Caldwell, C. N., 1980, ApJ,
237, 692
\reference{Lauer} Lauer, T. R. 2012, arXiv:1209.4357
\reference{Lau}Lauer, T. et al., 2002, AJ, 124, 1975
\reference{Lauer95} Lauer, T. R., Ajhar, E. A., Byun, Y.-I., et al. 1995, AJ, 110, 2622
\reference{laer3}Lauer, T. R., et al.\ 2005, AJ, 129, 2138
\reference{lau9}Lauer, T. R., et al., 2007a, ApJ, 662, 808
\reference{lawq}Lauer, T. R., et al., 2007b, ApJ, 664, 226
\reference{laurk}Laurikainen, E., Salo H., Buta R., Knapen J. H., \& Comer\'on S. 2010, MNRAS, 405, 1089
\reference{Li}Li, Z.-Y., Ho, L. C., Barth, A. J., \& Peng, C. Y. 2011, ApJS, 197,
22
\reference{longo} Longo, G., Zaggia, S. R., Busarello, G., \& Richter, G. 1994, A\&AS, 105, 433
\reference{MGC09}MacArthur, L.A., Gonz\'alez, J.J., Courteau, S.\ 2009, MNRAS, 395, 28

\reference{magor} Magorrian, J., et al. 1998, AJ, 115, 2285
\reference{man} Man, A. W. S., Toft, S., Zirm, A. W., Wuyts, S., van der Wel, A., 2012, ApJ,
744, 85
\reference{Mart} Martizzi, D., Teyssier R., \& Moore B. 2012a, MNRAS, 420, 2859
\reference{Marti} Martizzi, D., Teyssier R., \& Moore B., 2012b, arXiv:1211.2648
\reference{Meert} Meert, A., Vikram, V., \& Bernardi, M. 2012, arXiv:1211.6123
\reference{Merluzzi} Merluzzi, P., Busarello, G., Dopita, M.A., et al.\ 2012, MNRAS, arXiv:1211.6532
\reference{mikrr} Merritt, D., 2006, ApJ, 648, 976
\reference{MIch} Michard, R., \& Marchal, J. 1993, A\&AS, 98, 29
\reference{mikall} Milosavljevi\'c, M., \& Merritt, D. 2001, ApJ, 563, 34
\reference{milonMer}Milosavljevi\'c,  M., Merritt, D., Rest A., \& van den Bosch, F. C. 2002, MNRAS,
331, L51

\reference{moor} Moore, B., Katz, N., Lake, G., Dressler, A., Oemler, A., 1996, Nature,
379, 613
\reference{mulch}Mulchaey, J. S., Davis, D. S., Mushotzky, R. F., \& Burstein, D.
1993, ApJ, 404, L9
\reference{nieto} Nieto, J.-P., \& Bender, R. 1989, A\&A, 215, 266
\reference{Nipo} Nipoti, C., Londrillo, P., Ciotti, L., 2006, MNRAS, 370, 681
\reference{okam}Okamoto, T., Nagashima, M., 2001, ApJ, 547, 109
\reference{pat} Paturel, G., Petit, C., Prugniel, P., Theureau, G., Rousseau, J., Brouty, M.,
Dubois P., Cambr\'esy, L., 2003, A\&A, 412, 45
\reference{Pel90}Peletier, R.F., Davies, R.L., Illingworth, G.D., Davis, L.E., Cawson, M. 1990, AJ, 100, 1091
 \reference{Peirani}Peirani, S., Kay, S., \& Silk, J.\ 2008, A\&A, 479, 123
\reference{pic}Pichon, C., Pogosyan, D., Kimm, T., et al. 2011, MNRAS, 418, 2493
\reference{pink} Pinkney, J., Gebhardt, K., Bender, R., et al. 2003, ApJ, 596, 903
\reference{poggia} Poggianti, B. M., et al., 2009, ApJL, 697, 137
\reference{poggian} Poggianti, B. M., et al., ApJ, arXiv:1211.1005 
\reference{post}Postman, M., et al., 2012, ApJ, arXiv:1205.3839
 \reference{postman}Postman, M., \& Geller, M. J. 1984, ApJ, 281, 95
\reference{Quil}Quilis, V., Moore, B., \& Bower, R. 2000, Science, 288, 1617
\reference{Rav02}Ravindranath, S., Ho, L. C., \& Filippenko, A. V. 2002, ApJ, 566,
801
\reference{ravi}Ravindranath, S., Ho, L. C., Peng, C. Y., Filippenko, A. V., \& Sargent, W. L. W.\ 2001. AJ, 122, 653
\reference{reda}Reda, F. M., Forbes, D. A., Beasley, M. A., O’Sullivan, E. J., Goudfrooij, P., 2004, MNRAS, 354, 851
\reference{rest}Rest, A., van den Bosch, F. C., Jaffe, W., Tran, H., Tsvetanov, Z., Ford, H. C.,
Davies, J., \& Schafer, J.\ 2001, AJ, 121, 2431
\reference{riching} Richings, A. J., Uttley, P., \& Kr$\ddot{\textrm{o}}$ding, E., 2011, MNRAS, tmp, 759
\reference{richsto} Richstone, D., et al. 1998, Nature, 395, 14
\reference{salesl}Sales, L. V., Navarro, J. F., Theuns, T., Schaye, J., White, S. D. M.,
Frenk C. S., Crain, R. A., Dalla Vecchia, C., 2012, MNRAS,
423, 1544
\reference{Sara1} Saracco, P., Longhetti, M., \& Gargiulo, A., 2011, MNRAS, 412, 2707
\reference{Sara2} Saraiva, M. F., Ferrari, F., Pastoriza, M. G., 1999, A\&A, 350, 339
\reference{schw} Schweizer, F., \& Seitzer, P. 1992, AJ, 104, 1039
\reference{sersic1} S\'ersic, J. L.\ 1963, Bolet\'in de la Asociaci\'on Argentina de Astronom\'ia, 6, 41
\reference{Ses}Sesana, A., 2010, ApJ, 719, 851
\reference{Sil}Sil’chenko, O. K., Moiseev A. V., \& Shulga A. P. 2010, AJ, 140, 1462
\reference{siri}Sirianni, M., Jee, M. J., Ben\'itez, N., et al. 2005, PASP, 117, 1049
\reference{stein}Steinmetz, M., Navarro, J. F., 2002, New Astronomy, 7, 155
\reference{SZomo} Szomoru, D., Franx, M., \& van Dokkum, P. G. 2012, ApJ, 749, 121
\reference{Tera}Terashima, Y., Ho, L. C., \& Ptak, A. F. 2000, ApJ, 539, 161
\reference{tor} Tonry, J. L., et al., 2001, ApJ, 546, 681
\reference{tooo} Toomre, A., \& Toomre, J. 1972, ApJ, 178, 623
\reference{Truconfpre} Trujillo, I. 2012, arXiv:1211.3771
\reference{trui} Trujillo, I., Erwin, P., Asensio Ramos, A., \& Graham, A. W.\ 2004, AJ, 127,
1917
\reference{Trcompc} Trujillo, I., Feulner, G., Goranova, Y., et al. 2006, MNRAS, 373, L36
\reference{Turner} Turner, M. L., C\^ot\'e, P., Ferrarese, L., et al., arXiv:1208.0338 
\reference{vnd} van den Bergh, S., 2009, ApJ, 702, 1502
\reference{vandn} van den Bergh, S., 1982, PASP, 94, 459
\reference{vanwel}van der Wel, A., Rix H.-W., \& Wuyts S. et al.\ 2011, ApJ, 730, 38
\reference{voll}Vollmer, B., Braine, J., Pappalardo, C., \& Hily-Blant, P. 2008b,
A\&A, 491, 455 
\reference{voll}Vollmer, B., Soida, M., Chung, A., et al. 2008a, A\&A, 483, 89
\reference{wilm}Wilman, D. J., Oemler, A., Mulchaey, J. S., McGee S. L.,
Balogh M. L., \& Bower, R. G. 2009, ApJ, 692, 298
\reference{white}White, S. D. M., Rees, M. J., 1978, MNRAS, 183, 34
\reference{Xu011}Xu, C. K., Zhao, Y., Scoville, N., et al. 2012, ApJ, 747, 85
\reference{yu}Yu, Q., 2002, MNRAS, 331, 935
\end{references}
\end{document}